 \documentclass[twocolumn,pra,superscriptaddress,showpacs,floatfix]{revtex4}
\usepackage{amsmath}
\usepackage{soul}
\usepackage{color}
\usepackage{graphicx}
\usepackage{adjustbox}
\graphicspath{.}

\begin{document}

%
\title{Electric dipole moments of superheavy elements \\
  {A case study on copernicium}}

\author{Laima Rad\v{z}i\={u}t\.{e}}
\author{Gediminas Gaigalas}
\affiliation{Vilnius University,
             Institute of Theoretical Physics and Astronomy, 
	     Saul\.{e}tekio~al.~3, LT-10222, Vilnius, Lithuania}
%
\author{Per J{\"o}nsson}
\affiliation{Group for Materials Science and Applied Mathematics,
             Malm\"o University, S-20506, Malm\"o, Sweden}
\author{Jacek Biero{\'n}}
\email[]{Jacek.Bieron@uj.edu.pl}
\affiliation{Instytut Fizyki imienia~Mariana~Smoluchowskiego,
             Uniwersytet Jagiello{\'n}ski,
             Krak{\'o}w, Poland }

\date{\today}

\begin{abstract}
The multiconfiguration Dirac-Hartree-Fock (MCDHF) method {was} employed
to calculate atomic electric dipole moments (EDM)
of the superheavy element copernicium {(Cn, $Z=112$).}
The EDM {enhancement factors of Cn, here calculated for the first time,}
are about {one} order of magnitude larger than those of Hg.
The exponential dependence of {the} enhancement factors on {the} atomic number $Z$
along          group 12 of the periodic table was derived from
the EDMs of the entire {homologous} series,
 $^{69}_{30}$Zn,
$^{111}_{\phantom{1}48}$Cd,
$^{199}_{\phantom{1}80}$Hg,
$^{285}_{112}$Cn, and 
$^{482}_{162}$Uhb.
%
These results {show} that superheavy elements with sufficiently
{long}
half-lives {are} potential candidates for EDM searches.
\end{abstract}


\maketitle














\section{Introduction}
\label{section.Introduction}

The existence of a non-zero permanent electric dipole moment
(EDM) of an elementary {particle,} or in a nondegenerate system
of {particles,}  would be one {manifestation} of
violation of parity ($P$) and time reversal ($T$)
symmetries~\cite{KhriplovichLamoreaux:1997,Jungmann-2013}.
Violation of $P$ symmetry has been observed 
in the $\beta$-decay of $^{60}$Co~\cite{Wu:1957}
followed by decay of
muons~\cite{Garwin:muon:1957}
and
pions~\cite{Friedman:pion:1957}.
%
Violation of 
{charge and parity} ($CP$) symmetry has been observed
in the weak decay of neutral kaons K$^0${~\cite{Christenson:kaon:1957}}.
Violation of {$T$} symmetry is in turn equivalent
to violation of combined $CP$ symmetry, through the combined $CPT$
symmetry, which is considered invariant~\cite{Kostelecky-2011}.
%
%
 Both $CP$ and $T$ symmetry violations have been observed
 in the neutral kaon
 system~\cite{CPLEAR:1998},
 although direct $T$ symmetry violation has been
 disputed~\cite{Wolfenstein:1999,Bernabeu:2013}.
%
More recently a direct observation of the
$T$ symmetry violation 
in the {$B$} meson system has been reported~\cite{BABAR:2012}.
%
The violations of $P$, $C$, $CP$, and $T$ symmetries
are predicted by the Standard Model (SM) of
particle physics~\cite{Cabibbo:1963,Sozzi:book:2008}.
However, {the SM leaves several issues unexplained}, such as
the origin of baryogenesis,
the mass hierarchy of fundamental particles,
{the} number of particle generations,
{the} matter-antimatter-asymmetry observed in the universe, {and}
the nature of the dark matter. 
These {and other} issues are addressed within a large number of 
extensions of the present version of the {SM}.
Several of these extensions predict
EDMs induced by the $P$ and $T$ violating interactions
{and also}  EDMs of the fundamental particles
significantly larger than the values predicted by the {SM} itself.
The predictions can be tested, and 
searches for permanent electric dipole moment
are underway presently in various systems --- 
neutrons~\cite{Baker:2006},
{electrons in para- and diamagnetic atoms~\cite{Regan:2002,Griffith:2009}},
molecules~\cite{Hinds:2011,Baron:2014},
and other species~\cite{KhriplovichLamoreaux:1997,%
Ginges:2004,RobertsMarciano:2009}.
{The experimental searches have not yet detected a non-zero EDM,
but they continue to improve the limits on EDMs of individual
elementary particles, as well as limits on {$CP$}-violating
interactions, parametrized by the coupling constants
$C_{T}$ and
$C_{P}$
(see
references~\cite{KhriplovichLamoreaux:1997,%
Dzuba-Flambaum-Harabati-ratios:2011,%
GouldMunger:2014}
%
for details,
and the Table~II in the reference~\cite{Swallows:2013} for a summary).}

The primary objective of the present paper is the calculation of EDM
for {the} superheavy element
copernicium~\cite{Cn-112:Nature:2007,Cn-112:IUPAC:2009}.
We evaluated the contributions
to the atomic EDM
induced by four mechanisms~\cite{Ginges:2004}:
tensor-pseudotensor (TPT) and
pseudoscalar-scalar (PSS) interactions,
nuclear Schiff moment (NSM),
and 
electron electric dipole moment interaction with the nuclear magnetic
field (eEDM).
In each case we show {that there is} an order of magnitude increase of atomic EDM
between mercury and copernicium.
The second objective of the present paper is
{to derive the $Z$}-dependence of atomic EDM. 
We show that numerical EDM results
are consistent with {an} exponential {$Z$}-dependence
along the group 12 elements.

%
\section{MCDHF theory}
\label{sectionMCDHFtheory} 

We used the MCDHF approach to generate numerical representations
of atomic wave functions.
An atomic state function
$\Psi (\gamma P J M_{J})$
 is obtained as a linear combination of configuration state functions  
$\Phi (\gamma_{r} P J M_{J})$ {that are}
eigenfunctions of the parity $P$,
and total angular momentum operators $J^2$ and $M_{J}$:
%
\begin{equation}
\label{ASF}
\Psi (\gamma P J M_J) = \sum_{r} c_{r} \Phi (\gamma_{r} P J M_{J}).
\end{equation}
%
The multiconfiguration energy functional was based on the 
Dirac-Coulomb Hamiltonian, given (in a.u.) by
\begin{equation}
\label{eg:MCDF}
\hat{H}_{DC}  =
\sum_{j=1}^N
\Big( c \boldsymbol \alpha_j \cdot \boldsymbol  p_j +
 (\beta_j-1)c^2 +V({r_j})\Big)
 + \sum_{j < k}^N
\frac{1}{r_{jk}},
\end{equation}
where $\boldsymbol \alpha$ and $\beta$ are the Dirac matrices,
and $\boldsymbol p$ is the momentum operator.
The electrostatic electron-nucleus interaction,
$V({r_j})$,
{was} generated from  a 
nuclear charge density distribution
$\rho \left( r \right)$, which {was} approximated {by} the
normalized to $Z$ two-component
Fermi function~\cite{grasp89}
\begin{equation}
\label{eq:fermi2}
\rho(r)=\frac{\rho_0}{1+e^{(r-b)/a}},
\end{equation}
where $a$ and $b$ depend on the mass of the isotope.
%
%
The effects of the Breit interaction, as well as QED effects, were
neglected, since they are expected to be small at the level
of accuracy attainable in the present calculations.

%
\section{MCDHF wave functions}
\label{section.MCDHFwavefunctions}
%

We calculated the wave functions
of five diamagnetic atoms of group 12,
and subsequently
the EDMs
{in the ground states}
of the entire {homologous} series,
$^{69}_{30}$Zn,
$^{111}_{\phantom{1}48}$Cd,
$^{199}_{\phantom{1}80}$Hg,
$^{285}_{112}$Cn, and 
{$^{482}_{162}$Uhb}.
The numerical representations of {the} wave functions
were generated with
the relativistic atomic structure package
GRASP2K~\cite{grasp2K:2013},
based on the multiconfiguration Dirac-Hartree-Fock (MCDHF)
method~\cite{graspMcKenzie1980,grasp89,grasp92,Grant1994,grasp2K,%
GrantBook2007}.
%
%
Electron correlation effects were evaluated with methods
described in our previous
papers~\cite{RadziuteRaHgYbEDM2014,BieronAu2009,Bieron:e-N:2015,Radziute2015}.
Core-valence and valence-valence correlations were included {by allowing} 
single and restricted double substitutions
to five sets of virtual orbitals.
%
%
{The full description of numerical methods, virtual orbital sets,
electron substitutions,
and other details of the computations can be found
in~\cite{RadziuteRaHgYbEDM2014}.
Compared with~\cite{RadziuteRaHgYbEDM2014} 
the double electron substitutions were however extended 
from the $nsnp$ to the $(n-1)dnsnp$ shells
in the present paper
(see section~\ref{section.Mercury} below for details).}

%
\section{EDM calculations}
\label{sectionEDMcalculations}
%
An atomic EDM can be written as {a} sum over states
(equation~(4) in reference~\cite{RadziuteRaHgYbEDM2014}):
\begin{eqnarray}
\label{eq:DAHint}
   d^{int}  = 
   \; 2  \sum_{i}
    \frac{ \left< 0| \hat{D}_{z} |i\right> 
   \left< i | \hat{H}_{int}    |0 \right>}
   {E_{0} \; - \; E_{i}} 
  ,
\end{eqnarray}
 where $ | 0  \rangle $ represents the ground
 state $|\Psi ( \gamma  P J M_J {)} \rangle$
 of a closed-shell atom from the group 12,
with $J=0$ and even parity. {The} summation runs over excited
states $|\Psi ( \gamma_{i} (-P) J_{i} M_{J_{i}} {)} \rangle$,
with $J_{i}=1$ and odd parity.
A calculation of an atomic EDM requires evaluation of the
matrix {elements} of the static dipole
$ \hat{D}_{z} $,
and the matrix elements  of the $ \hat{H}_{int} $
{interactions}, which {induce an} EDM in an atom~\cite{Gaigalas1997}.
In order to perform these calculations the GRASP2K
{package was} extended. 
The extension includes programs for matrix element {calculations}, 
based on spin-angular integration~\cite{Gaigalas1997}.
Here $ \hat{H}_{int} $
represents one of the four interactions mentioned 
in the section~\ref{section.Introduction} above,
$E_{0}$ and $E_{i}$
are energies  of ground and excited  states, respectively.
{The full description of the EDM theory underlying the presents
 calculations can be found in 
reference~\cite{RadziuteRaHgYbEDM2014}, and in references therein.}

%
%
{The summation in equation~(\ref{eq:DAHint}) involves {an} infinite number
of bound states, as well as {contributions} from the continuum
spectrum.
The sum over the bound spectrum was evaluated {by explicitly calculating
contributions from the lowest five odd states of each symmetry using
numerical wave functions. Then the method of 'Riemann zeta tail', described
in reference~\cite{RadziuteRaHgYbEDM2014}, was applied to sum up
the contribution from the remaining bound states. To this end
we showed that a summation over a Rydberg series, when extrapolated to
large values of the principal quantum number $n$ of the running electron
(and where the energy denominator saturates at the ionisation energy)
converges to the
Riemann zeta function.
The explicit numerical summation accounts for 98 percent of the
whole sum, and we evaluated the upper bound on the rest (the infinite tail)
of the sum by exploiting regularities of the Rydberg series.}
The relative  correction,
i.e.~the total contribution from the trailing terms
(called Riemann zeta tail)
divided by the total contribution from the five leading terms,
is of the order of 1.5 percent for mercury $^{199}$Hg, and
below 2 percent for copernicium $^{285}$Cn.
We neglected the Riemann zeta tail correction for the other
three elements ($^{69}$Zn, $^{111}$Cd, $^{482}$Uhb).}

%
%
{The contribution from the continuum is difficult to estimate,
since it is partially accounted for by the virtual
set~\cite{CowanHansen:1981}
In the present paper
we computed only the contribution of the bound states.
{We} neglected the explicit summation over continuum,
and assumed that the continuum spectrum contribution
were included into the error budget.
The evaluation of the sum over the continuum part of the spectrum could
in principle be carried out either fully
numerically~\cite{edm-continuum-work-in-progress},
or again by an
extrapolation, based on the fact, that the oscillator strength density is
continuous across the ionization threshold~\cite{FanoCooper:1965},
and above mentioned regularities carry over to the continuum spectrum.}

%
%
{The electronic matrix elements 
in equation~(\ref{eq:DAHint}) are not isotope-specific.
However, the atomic {wave functions} do exhibit a (rather weak) dependence
on the atomic mass of the element of interest, through the nuclear
electrostatic potential, which depends on 
the nuclear charge density distribution, which in turn depends on 
the nuclear mass number, through the equation~(\ref{eq:fermi2}).
Therefore, all numerical results in 
Tables~\ref{table_1_TPT},
\ref{table_2_PSS},
\ref{table_3_NSM},
and~\ref{table_4_eEDM}
were obtained for specific isotopes, such as $^{199}$Hg and $^{285}$Cn,
and they do exhibit a (negligibly weak) dependence on atomic masses.}
%
%
%
\begin{table}[htbp]                                                             
\begin{scriptsize}  
\tabcolsep=0.045cm                                                            
\caption{TPT interaction contributions to EDM 
in different virtual sets,                       
in units                                                                        
 $\left(10^{-20} C_T \left< {\bf \sigma}_{A} \right> \left|e\right|             
  \mbox{cm} \right)$,                                                           
for $^{69}$Zn, $^{111}$Cd, $^{199}$Hg, and $^{285}$Cn, 
compared with data from other methods.
 See text for explanations and details.
 }
\label{table_1_TPT}                                                   
\begin{tabular}{lllllllllllll}  
\hline\noalign{\smallskip}                                                      
\hline\noalign{\smallskip}                                                      
 &\multicolumn{2}{c}{$^{69}$Zn }    
&&\multicolumn{2}{c}{$^{111}$Cd}
&&\multicolumn{2}{c}{$^{199}$Hg}  
&&\multicolumn{3}{c}{$^{285}$Cn}  \\
\cline{2-3}\cline{5-6}\cline{8-9}\cline{11-13} 
\multicolumn{1}{c}{VO\rlap{S}} &
\multicolumn{1}{c}{\phantom{-}Th} &
\multicolumn{1}{c}{\phantom{-}SE} &&
\multicolumn{1}{c}{\phantom{-}Th} &
\multicolumn{1}{c}{\phantom{-}SE} &&
\multicolumn{1}{c}{\phantom{-}Th} &
\multicolumn{1}{c}{\phantom{-}SE} &&
\multicolumn{1}{c}{\phantom{1}Th} &
\multicolumn{1}{c}{\phantom{-}Th2} &
\multicolumn{1}{c}{\phantom{-}Th3} \\
\hline                                                                          
\phantom{1}DF & $-$0.07 & $-$0.07&&$-$0.35 & $-$0.36&& $-$7.29 &$-$6.15  && $-$59.86 & $-$61.50 & $-$66.66\\
\phantom{1}1  & $-$0.08 & $-$0.09&&$-$0.39 & $-$0.45&& $-$4.13 &$-$4.86  && $-$48.53 & $-$50.95 & $-$53.95\\
\phantom{1}2  & $-$0.09 & $-$0.11&&$-$0.45 & $-$0.54&& $-$4.66 &$-$5.23  && $-$58.38 & $-$58.92 & $-$62.96\\
\phantom{1}3  & $-$0.10 & $-$0.12&&$-$0.47 & $-$0.57&& $-$4.84 &$-$5.53  && $-$59.31 & $-$64.53 & $-$68.76\\
\phantom{1}4  & $-$0.10 & $-$0.12&&$-$0.48 & $-$0.59&& $-$4.79 &$-$5.64  && $-$57.67 & $-$61.04 & $-$65.26\\
\phantom{1}5  & $-$0.11 & $-$0.12&&$-$0.49 & $-$0.60&& $-$4.84 &$-$5.64  && $-$57.51 & $-$60.75 & $-$64.98\\
\hline                                                                          
\multicolumn{5}{l}{Ref.~\cite{DzubaFlambaum:2009}(DHF)    } &    &&&$-$2.4  \\
\multicolumn{5}{l}{Ref.~\cite{Martensson:1985}(DHF)       } &    &&&$-$2.0  \\
\multicolumn{5}{l}{Ref.~\cite{DzubaFlambaum:2009}(CI+MBPT)} &    &&&$-$5.12 \\
\multicolumn{5}{l}{Ref.~\cite{DzubaFlambaum:2009}(RPA)    } &    &&&$-$5.89 \\
\multicolumn{5}{l}{Ref.~\cite{Martensson:1985}(RPA)       } &    &&&$-$6.0  \\
\multicolumn{5}{l}{Ref.~\cite{Latha:2008}(CPHF)           } &    &&&$-$6.75 \\
\multicolumn{5}{l}{Ref.~\cite{Latha:2009}(CCSD)           } &    &&&$-$4.3  \\
\hline                                                                          
%
\end{tabular} 
\end{scriptsize}   
\end{table}
%
%
\section{Mercury}
\label{section.Mercury}
%
The calculations for $^{199}$Hg were performed in a similar manner
as those presented in our previous paper~\cite{RadziuteRaHgYbEDM2014}.
The {results from DF} and {from calculations with the} first two layers
of virtual orbitals
(i.e.~the first three lines in 
Tables~\ref{table_1_TPT},
\ref{table_2_PSS},
\ref{table_3_NSM},
and~\ref{table_4_eEDM})
are in fact identical {with the results published
in~\cite{RadziuteRaHgYbEDM2014}}.
Further calculations differ in the scope of the double electron substitutions,
which were extended from $6s6p$ to  $5d6s6p$ shells.
%
\begin{table}[htbp]                                                             
\begin{scriptsize} 
\tabcolsep=0.01cm                                                               
\caption{PSS interaction contributions to EDM 
in different virtual sets,                       
in units                                                                        
 $\left( 10^{-23} C_P \left< \sigma_{A} \right> \left| e \right|
 \mbox{cm} \right)$,                                                           
for $^{69}$Zn, $^{111}$Cd, $^{199}$Hg, and $^{285}$Cn, 
compared with data from other methods.
 See text for explanations and details.
 }                                         
\label{table_2_PSS}                                                   
\begin{tabular}{lllllllllllll} 
\hline\noalign{\smallskip}  
\hline\noalign{\smallskip}  
 &\multicolumn{2}{c}{$^{69}$Zn } 
&&\multicolumn{2}{c}{$^{111}$Cd}
&&\multicolumn{2}{c}{$^{199}$Hg}   
&&\multicolumn{3}{c}{$^{285}$Cn}  \\
\cline{2-3}\cline{5-6}\cline{8-9}\cline{11-13} 
\multicolumn{1}{c}{VO\rlap{S}} &
\multicolumn{1}{c}{\phantom{-}Th} &
\multicolumn{1}{c}{\phantom{-}SE} &&
\multicolumn{1}{c}{\phantom{-}Th} &
\multicolumn{1}{c}{\phantom{-}SE} &&
\multicolumn{1}{c}{\phantom{-}Th} &
\multicolumn{1}{c}{\phantom{-}SE} &&
\multicolumn{1}{c}{\phantom{1}Th} &
\multicolumn{1}{c}{\phantom{-}Th2} &
\multicolumn{1}{c}{\phantom{-}Th3} \\
\hline                                                                          
\phantom{1}DF                          &$-$0.13& $-$0.14&& $-$0.94& $-$0.96 &&$-$25.47 &$-$21.49&&$-$199.52&$-$252.66& $-$274.11\\                   
\phantom{1}1                           &$-$0.15& $-$0.17&& $-$1.05& $-$1.21 &&$-$14.54 &$-$17.16&&$-$199.52&$-$209.13& $-$221.73\\                   
\phantom{1}2                           &$-$0.19& $-$0.23&& $-$1.19& $-$1.46 &&$-$16.38 &$-$18.39&&$-$240.22&$-$242.15& $-$259.07\\                   
\phantom{1}3                           &$-$0.20& $-$0.24&& $-$1.25& $-$1.53 &&$-$17.01 &$-$19.47&&$-$244.96&$-$266.95& $-$284.65\\                   
\phantom{1}4                           &$-$0.20& $-$0.24&& $-$1.28& $-$1.58 &&$-$16.84 &$-$19.84&&$-$237.56&$-$251.33& $-$268.95\\                   
\phantom{1}5                           &$-$0.22& $-$0.24&& $-$1.30& $-$1.60 &&$-$17.02 &$-$19.85&&$-$236.88&$-$250.07& $-$267.78\\                   
\hline 
\multicolumn{5}{l}{Ref.~\cite{DzubaFlambaum:2009}(DHF)      }&    &&&~\,$-$8.7   &\\
\multicolumn{5}{l}{Ref.~\cite{DzubaFlambaum:2009}(CI+MBPT)  }&    &&&$-$18.4  &\\
\multicolumn{5}{l}{Ref.~\cite{DzubaFlambaum:2009}(RPA)      }&    &&&$-$20.7  &\\
\hline
\end{tabular}                                                                   
\end{scriptsize}                                                                
\end{table}                                                                     
%
%
\begin{table}[htbp]                                                             
\begin{scriptsize}  
\tabcolsep=0.045cm                                                            
\caption{Schiff moment contributions to atomic EDM in different virtual sets,
 in units 
 $\left\{ 10^{-17} [S/(\left| e \right| \mbox{fm}^3)] \left| e \right|
 \mbox{cm} \right\}$,                                                           
for $^{69}$Zn, $^{111}$Cd, $^{199}$Hg, and $^{285}$Cn, 
compared with data from other methods.
 See text for explanations and details.
 }                                         
\label{table_3_NSM}                                                   
\begin{tabular}{lllllllllllll} 
\hline\noalign{\smallskip} 
\hline\noalign{\smallskip} 
  &\multicolumn{2}{c}{$^{69}$Zn } 
&&\multicolumn{2}{c}{$^{111}$Cd}
&&\multicolumn{2}{c}{$^{199}$Hg} 
&&\multicolumn{3}{c}{$^{285}$Cn}  \\
\cline{2-3}\cline{5-6}\cline{8-9}\cline{11-13} 
\multicolumn{1}{c}{VO\rlap{S}} &
\multicolumn{1}{c}{\phantom{-}Th} &
\multicolumn{1}{c}{\phantom{-}SE} &&
\multicolumn{1}{c}{\phantom{-}Th} &
\multicolumn{1}{c}{\phantom{-}SE} &&
\multicolumn{1}{c}{\phantom{-}Th} &
\multicolumn{1}{c}{\phantom{-}SE} &&
\multicolumn{1}{c}{\phantom{1}Th} &
\multicolumn{1}{c}{\phantom{-}Th2} &
\multicolumn{1}{c}{\phantom{-}Th3} \\
\hline                                                                          
\phantom{1}DF                     &$-$0.04 & $-$0.04 &&$-$0.18  & $-$0.19 &&$-$2.86 &$-$2.46&&$-$17.73&$-$17.26&$-$19.53\\
\phantom{1}1                      &$-$0.05 & $-$0.06 &&$-$0.21  & $-$0.26 &&$-$1.95 &$-$2.45&&$-$13.64&$-$12.96&$-$14.53\\
\phantom{1}2                      &$-$0.06 & $-$0.07 &&$-$0.25  & $-$0.32 &&$-$2.11 &$-$2.42&&$-$17.05&$-$15.96&$-$17.78\\
\phantom{1}3                      &$-$0.06 & $-$0.08 &&$-$0.27  & $-$0.34 &&$-$2.21 &$-$2.58&&$-$20.09&$-$22.66&$-$24.58\\
\phantom{1}4                      &$-$0.06 & $-$0.08 &&$-$0.28  & $-$0.35 &&$-$2.19 &$-$2.62&&$-$17.75&$-$18.02&$-$19.95\\
\phantom{1}5                      &$-$0.07 & $-$0.08 &&$-$0.28  & $-$0.35 &&$-$2.22 &$-$2.63&&$-$17.62&$-$17.77&$-$19.71\\        
\hline                                                                          
\multicolumn{5}{l}{Ref.~\cite{DzubaFlambaum:2009}(DHF)     }&& &&$-$1.2   \\
\multicolumn{5}{l}{Ref.~\cite{DzubaFlambaum:2009}(CI+MBPT) }&& &&$-$2.63  \\
\multicolumn{5}{l}{Ref.~\cite{DzubaFlambaum:2009}(RPA)     }&& &&$-$2.99  \\
\multicolumn{5}{l}{Ref.~\cite{Dzuba:2002}(CI+MBPT)         }&& &&$-$2.8   \\
\multicolumn{5}{l}{Ref.~\cite{DzubaFlambaum:2007:76}(TDHF) }&& &&$-$2.97  \\
\multicolumn{5}{l}{Ref.~\cite{Latha:2009}(CCSD)            }&& &&$-$5.07  \\
\hline                                                                          
\end{tabular}                                                                   
\end{scriptsize}                                                                
\end{table}
%
%
%
\begin{table}[htbp]                                                             
\begin{scriptsize}  
\tabcolsep=0.06cm                                                             
\caption{Contributions of electron EDM interaction with magnetic field
of nucleus,
to atomic EDM in different virtual sets,
in units ($d_e \times 10^{-4}$),
for $^{69}$Zn, $^{111}$Cd, $^{199}$Hg, and $^{285}$Cn, 
compared with data from other methods.
 See text for explanations and details.
 }                                         
\label{table_4_eEDM}                                                   
\begin{tabular}{lllllllllllll} 
\hline\noalign{\smallskip} 
\hline\noalign{\smallskip}  
 &\multicolumn{2}{c}{$^{69}$Zn }    
&&\multicolumn{2}{c}{$^{111}$Cd}
&&\multicolumn{2}{c}{$^{199}$Hg} 
&&\multicolumn{3}{c}{$^{285}$Cn}  \\
\cline{2-3}\cline{5-6}\cline{8-9}\cline{11-13} 
\multicolumn{1}{c}{VO\rlap{S}} &
\multicolumn{1}{c}{\phantom{-}Th} &
\multicolumn{1}{c}{\phantom{-}SE} &&
\multicolumn{1}{c}{\phantom{-}Th} &
\multicolumn{1}{c}{\phantom{-}SE} &&
\multicolumn{1}{c}{\phantom{-}Th} &
\multicolumn{1}{c}{\phantom{-}SE} &&
\multicolumn{1}{c}{\phantom{1}Th} &
\multicolumn{1}{c}{\phantom{-}Th2} &
\multicolumn{1}{c}{\phantom{-}Th3} \\
\hline                                                                          
\phantom{1}DF                      & 0.13 & 0.14 &&$-$0.62 &$-$0.63 && 16.04 & 13.41  && 314.03 & 324.40 & 350.09 \\ 
\phantom{1}1                       & 0.11 & 0.09 &&$-$0.64 &$-$0.71 && 8.47  &  9.58  && 254.78 & 269.22 & 283.51 \\ 
\phantom{1}2                       & 0.13 & 0.13 &&$-$0.69 &$-$0.81 && 9.63  & 10.64  && 305.55 & 309.48 & 328.86 \\ 
\phantom{1}3                       & 0.14 & 0.14 &&$-$0.72 &$-$0.85 && 9.99  & 11.30  && 305.13 & 329.18 & 349.47 \\ 
\phantom{1}4                       & 0.14 & 0.14 &&$-$0.73 &$-$0.87 && 9.90  & 11.53  && 300.39 & 318.41 & 338.60 \\ 
\phantom{1}5                       & 0.13 & 0.11 &&$-$0.75 &$-$0.88 && 10.00 & 11.50  && 299.67 & 317.11 & 337.40 \\                   
\colrule                                                                     
\multicolumn{5}{l}{Ref.~\cite{DzubaFlambaum:2009}(DHF)    }&&&&  ~\,4.9   \\
\multicolumn{5}{l}{Ref.~\cite{Martensson:1987}(DHF)       }&&&&  ~\,5.1   \\
\multicolumn{5}{l}{Ref.~\cite{DzubaFlambaum:2009}(CI+MBPT)}&&&&  10.7  \\
\multicolumn{5}{l}{Ref.~\cite{DzubaFlambaum:2009}(RPA)    }&&&&  12.3  \\
\multicolumn{5}{l}{Ref.~\cite{Martensson:1987}(RPA)       }&&&&  13    \\
\hline                                                                          
\end{tabular}                                                                   
\end{scriptsize}                                                                
\end{table}                                                              
%
%
%
%
The results of the calculations are presented in
Tables~\ref{table_1_TPT},
\ref{table_2_PSS},
\ref{table_3_NSM},
and~\ref{table_4_eEDM}.
{The number of virtual orbital sets (VOS) is listed in the first column
of each table
%
%
(see chapter~III of reference~\cite{RadziuteRaHgYbEDM2014} for
definitions and for the details of the calculations).
The line marked 'DF' (Dirac-Fock) in the VOS column represents
the lowest-order approximation, with zero sets of virtual orbitals.
%
%
{It should be noted that the values in the Tables
 marked 'DF' and 'DHF' are not equivalent. Those marked 'DF'
were obtained in the present calculations with only
{the} two lowest excited states
included in the summation in equation~(\ref{eq:DAHint}).
The results marked 'DHF', obtained with MBPT methods, involved
summation over the entire spectrum of virtual orbitals,
using various methods to construct the virtual orbital 
set~\cite{Martensson:1985,Martensson:1987,DzubaFlambaum:2009}.
{Neither 'DF' nor 'DHF' include electron correlation effects and
therefore they are relevant only for the purpose of evaluating
the contributions of electron correlation for the expectation
values of interest.}
}
A larger number of VOS represents {in principle} a better
approximations of the wave function.
The line marked '5' in the VOS column represents the final
approximation, with five sets of virtual orbitals
(MCDHF-VOS.5, represented by red circles in Figure~\ref{linZ_lnEDM}).}
The difference between VOS.4 and VOS.5
may {(cautiously) be} taken as
an indication of accuracy.
For each element the calculated values of the energy denominators
in equation~(\ref{eq:DAHint})
were used to evaluate the atomic EDMs.
{These} fully theoretical EDM values are marked 'Th' in
Tables~\ref{table_1_TPT},
\ref{table_2_PSS},
\ref{table_3_NSM},
and~\ref{table_4_eEDM}.
Semiempirical EDM values (marked 'SE' in the Tables) were also evaluated
for $^{69}$Zn, $^{111}$Cd, and $^{199}$Hg, with
the energy denominators taken from the NIST database~\cite{NIST_ASD}.
Level identifications were made with the atomic state functions
transformed from 
$jj$-coupling to $LSJ$ coupling scheme, using the methods developed
in \cite{Transformation,LSJ2}.
%

\section{Copernicium}
\label{section.Copernicium}
%
%
Three different sets of  energy denominators for $^{285}$Cn were used.
Those from {the present} calculations are marked 'Th'.
For comparison purposes we computed also the EDMs with the
energy denominators {taken} from two other theoretical
papers~\cite{Fritzsche:2007:2,Dinh:Dzuba:2008}.
The results in column marked 'Th2' were obtained with the
energy denominators taken from~\cite{Fritzsche:2007:2},
who used {a} large-scale {MCDHF} method.
The authors of~\cite{Fritzsche:2007:2}
evaluated also the ionization limit of copernicium {and}
their calculated ionization energy was used 
in our evaluation of EDMs
for those levels which were
not reported in~\cite{Fritzsche:2007:2}.
The energy denominators in {the columns} marked 'Th3'
were taken from
reference~\cite{Dinh:Dzuba:2008}, where
the energy spectrum was computed with two
methods: CI+MBPT and CI.
We gave priority to the CI+MBPT results;
the CI results were used when CI+MBPT data were not available;
%
{for the remaining levels the energy denominators were replaced
 by the calculated ground state ionization energy.}
The accuracy of our calculated energy values, as well as those from the
references~\cite{Fritzsche:2007:2} and~\cite{Dinh:Dzuba:2008},
is better than 20\% for the lowest excited levels of mercury.
%
%

%
%
The mass number 285 for the element Cn was chosen 
due to predictions that heavier isotopes are more stable
than the
lighter ones~\cite{Smolanczuk:1997,Zagrebaev:2013}.
%
The lifetimes of several known isotopes of Cn
are counted in minutes~\cite{nudat2},
which make them amenable to atom traps,
and subsequent spectroscopy.
It is predicted that still heavier isotopes of {Cn},
with mass numbers in the range  290--294,
may have half-lives counted in years~\cite{Zagrebaev:2013}.
%

%
%
%
%
We observed a similar pattern of contributions from individual electronic
states, as described in~\cite{RadziuteRaHgYbEDM2014}.
The triplet $6snp$~$^3\!P_1$
 and the singlet $6snp$~$^1\!P_1$ states
 are the dominant contributors to atomic EDM in the $^{199}$Hg spectrum.
For the $^{285}$Cn case
the dominant contributions 
arise from the lowest states of $  ^{1,3}\!P_1 $ symmetries, 
i.e.~$ 7snp \, ^1\!P_1 $,
$ 7snp \, ^3\!P_1 $,
Altogether they contribute in excess of 98\% of the total EDM.
%
The remaining Rydberg states contribute less than 2 percent.
%
%
Instead of 
an explicit error analysis for the
calculations of EDM for  $^{285}$Cn
we applied a comparison with mercury.
Estimates of the magnitudes of EDMs
induced by {the} TPT, PSS, NSM, and eEDM mechanisms
in mercury, have been performed with several theoretical
methods~\cite{DzubaFlambaum:2007:76,
DzubaFlambaum:2009,
Latha:2008,
Latha:2009}. 
With one or two
exceptions~\cite{Latha:2009,Graner:2016},
they all agree within reasonable error bounds --- of the order of 10--20
percent~\cite{RadziuteRaHgYbEDM2014}.
The results of the MCDHF calculations for 
mercury, both in present paper as well as in~\cite{RadziuteRaHgYbEDM2014},
are well within these bounds.
We expect that the present calculations for $^{285}$Cn,
performed with the same MCDHF model as those for $^{199}$Hg,
would also fit within error bounds of similar size.

\section{Unhexbium}
\label{section.Unhexbium}
%
In addition to the calculations described above we have done
 uncorrelated DF calculations for $^{482}_{162}$Uhb and for $^{9}_{4}$Be.
%
%
There are several theoretical
predictions~\cite{Fricke:Z-172:1970,Penneman:Z-164:1971,Pyykko:Z-172:2011}
which suggest that the heaviest
{homologue} in the Zn-Cd-Hg-Cn-Uxx group would not be element E162
{(Unhexbium)},
but E164
(Unhexquadium).
Due to a very large spin-orbit splitting of the $8p$ shell,
the relativistic $8p_{1/2}$ shell becomes occupied before
the $7d$ shell is filled~\cite{Pyykko:Z-172:2011}.
Therefore, at the end of transition metals 
in the row eight of the periodic table appears the element E164,
with the ground configuration [Cn]$5g^{18} 6f^{14} 7d^{10} 7p^6 8s^2 8p^2 $,
with all inner shells closed,
and with two electrons in the $8p_{1/2}$ shell
(the $8p_{1/2}$ shell is, in fact, also closed).
However, the presence of the $8p$ shell would complicate the  calculations
of EDMs,
and, more importantly, {it}
would complicate comparisons along the {homologous} series. {Therefore}
we have deliberately chosen a {(doubly artificial)} isotope
$^{482}_{162}$Uhb,
of element E162,
with electron configuration 
[Cn]$5g^{18} 6f^{14} 7d^{10} 7p^6 8s^2 $.

\section{$Z$-dependence}
\label{section.Zdependence}

%
Atomic properties depend in various ways on the atomic number $Z$,
both in isoelectronic
{sequences~\cite{FBJbook,Layzer:1959,Layzer:1964,Edlen:1964},
as well as along homologous
sequences~\cite{Cowan.book,WieseWeiss:1968}.
In many cases approximate analytic relations were
derived~\cite{WieseWeiss:1968,Cowan.book,FBJbook,Edlen:1964},} 
and several atomic observables exhibit a polynomial or power dependencies 
on the atomic number $Z$.

%
Atomic enhancement factors of 
the $PT$-odd interactions in neutral atoms
scale with nuclear charge as
$d_{at} \sim \alpha^2 Z^3 $.
The factor $ Z^3 $ arises from an estimate of the strength
of the electric field in the vicinity of an atomic nucleus
(see chapter 6.2 of the reference~\cite{KhriplovichLamoreaux:1997}),
but it has been pointed out that  on top of the $ Z^3 $ enhancement
of the $PT$-odd interaction there is another
%
%
%
enhancement factor,
arising from relativistic contraction of the electronic
wave function~\cite{BouchiatBouchiat:1974,%
Bouchiat-Bouchiat-Phys.Lett.B48-111-1974.pdf,%
BouchiatBouchiat:1975,%
SushkovFlambaumKhriplovich:1984,%
Khriplovich:1991,KhriplovichLamoreaux:1997,%
FlambaumGinges:2002}.
%
%
\begin{equation}
\label{eq:K2gamma}
 K_{r} \approx 
        \left( \frac{\Gamma(3)}{\Gamma(2 \gamma + 1)}\right)^2
        \left( \frac{2 Z r_N}{a_0} \right) ^{2 \gamma -2}.
\end{equation}
%

%
{$Z$}-dependence of atomic EDMs induced by the $(P,T)$-odd $ \hat{H}_{int} $ 
interactions is governed by the {$Z$}-dependence of three factors
in equation~(\ref{eq:DAHint}):
matrix element of the $(P,T)$-odd $ \hat{H}_{int} $ operator,
matrix element of the electric dipole $ \hat{D}_{z} $ operator,
{and the} energy denominator
$ (E_{0} -  E_{i}) $.
The matrix elements of the electric dipole $ \hat{D}_{z} $ operator
are constrained by the Thomas-Reiche-Kuhn rule.
In case of the elements of group 12 they are further constrained
by the Wigner-Kirkwood sum rule
(see chapter 14 of the reference~\cite{Cowan.book}).
The two lines, 
$ ns^{2}   \,^1\!S_0 $~---~$ nsnp \, ^3\!P_1 $ and 
$ ns^{2}   \,^1\!S_0 $~---~$ nsnp \, ^1\!P_1 $,
dominate the Wigner-Kirkwood sum in all five elements,
making the
matrix element of $ \hat{D}_{z} $ approximately constant
along the homologous series.
Transition energy denominators in the equation~(\ref{eq:DAHint})
do not depend on {$Z$} along the homologous series~\cite{WieseWeiss:1968},
except {for} small variations due to shell contractions,
shell rearrangements, etc
(excluding the Uhb element,
with its large spin-orbit splitting mentioned 
in the section~\ref{section.Copernicium} above).

\begin{figure}
 \includegraphics[width=0.48\textwidth]{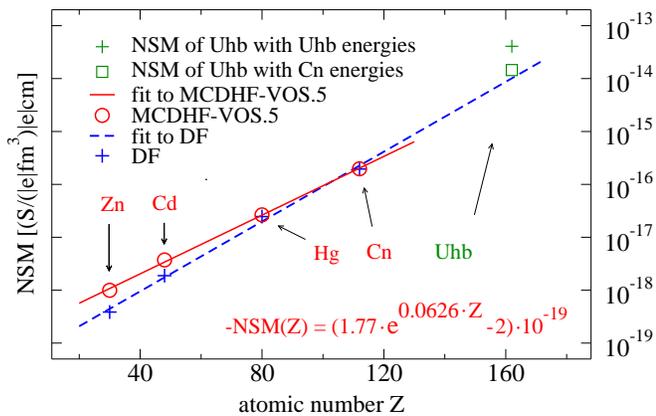}
\caption{(Color online): Atomic EDM (absolute values) induced by
 the NSM as a function of atomic number {$Z$}.
 Red {circles} = MCDHF-VOS.5 results with 5 virtual orbital sets.
 Blue {pluses} = uncorrelated DF results (0 sets).
 The lines are exponential functions fitted to the four points,
 representing Zn, Cd, Hg, and Cn.
 Solid red line = fit to MCDHF-VOS.5 results.
 Dashed blue line = fit to uncorrelated DF results.
 The lines are extrapolated beyond {$Z=112$}.
 The two symbols in the upper right corner
 represent Uhb (excluded from the fitting).
 Green plus = DF result for Uhb with calculated Uhb energy denominators.
 Green square = DF result for Uhb with energy denominators {taken} from Cn.
 The sizes of circles represent approximately the relative accuracy of the 
  MCDHF-VOS.5 calculations.
 See text for details.
 }
\label{linZ_lnEDM} 
\end{figure}

%
Therefore, the dominant role in
{establishing the $Z$-dependence of atomic EDMs along the
{homologous} sequence} is taken
by the $ \hat{H}_{int} $ operators.
%
%
{Following the analysis in chapter 8 of the
reference~\cite{KhriplovichLamoreaux:1997},
it can be shown that in the vicinity of a point-like atomic nucleus
the large $P_{n \kappa}$ and small $Q_{n \kappa}$ 
radial components of a one-electron {wave function} may be expressed as
%
%
\begin{equation}
\label{eq:ff}
P_{n \kappa} (r) = \frac{\kappa}{|\kappa|} (\kappa - \gamma)
     \left( \frac{Z}{a_0^3\nu^3} \right) ^{1/2}
     \frac{2}{\Gamma(2 \gamma + 1)}
     \left( \frac{a_0}{2 Z r} \right) ^{1-\gamma} \\
\end{equation}
\begin{equation}
\label{eq:gg}
Q_{n \kappa} (r) = \frac{\kappa}{|\kappa|} (Z \alpha)
     \left( \frac{Z}{a_0^3\nu^3} \right) ^{1/2}
     \frac{2}{\Gamma(2 \gamma + 1)}
     \left( \frac{a_0}{2 Z r} \right) ^{1-\gamma},
\end{equation}
where
$\kappa$ is the angular momentum quantum number,
$\gamma^2 = \kappa^2 - \alpha^2 Z^2$,
$\alpha$ is the fine structure constant,
$\nu^3$ is the effective principal quantum number,
{and} $a_0$ is {the} Bohr radius.
The radial integrals involved in the calculations of 
matrix elements of $ \hat{H}_{int} $ 
include the integrands of the combinations of 
the large $P_{n \kappa}$ and small $Q_{n \kappa}$ 
radial components, of the type 
$(P_a P_b \pm Q_a Q_b)$ 
or $(P_a Q_b \pm P_b Q_a)$.
All these integrals include factors in the integrands
which effectively cut off the integrals outside
the nucleus~\cite{RadziuteRaHgYbEDM2014},
and eventually
%
$Z$-dependence of the atomic EDM in the form
\begin{equation}
\label{eq:ffgg2gamma}
 d_{at} \sim 
        \left( \frac{ Z^k}{a_0^3\nu^3} \right)
        \left( \frac{2}{\Gamma(2 \gamma + 1)}\right)^2
        \left( \frac{2 Z r_N}{a_0} \right) ^{2 \gamma -2}
\end{equation}
is obtained, {where $k$ depends on a particular form of the integrand
and where $r_N$ is the effective cut off radius}.
{The
right hand side of the equation~(\ref{eq:ffgg2gamma})
is displayed in Fig.~\ref{f1g1}.
All four combinations
($P_a P_b $, $ Q_a Q_b$,
 $P_a Q_b $, and $ P_b Q_a$)
of the one-electron wave function factors
from equations~(\ref{eq:ff}) and~(\ref{eq:gg})
are represented
as functions of atomic number {$Z$}.
%
The index $a$ represents $ns_{1/2}$ orbitals,
the index $b$ represents $np_{3/2}$ orbitals.
The nuclear radius $r_N$ 
has been computed using $r_N = r_0 \cdot A^{1/3}$,
where $r_0 = 1.25$~fm.
The relation of atomic mass $A$ to atomic number $Z$ 
has been evaluated from the neutrons to protons ratio
$ N/Z = 1 + A^{2/3} a_C/2a_A $,
derived from the Bethe-Weizs{\"a}cker
formula~\cite{Rohlf:1994},
with
$ a_C = 0.711 $
and
$ a_A = 23.7  $.
The empty circles in the Fig.~\ref{f1g1}
show positions of
the four elements (Zn, Cd, Hg, Cn) considered in this paper.
} 
%
%
\begin{figure}
 \includegraphics[width=0.48\textwidth]{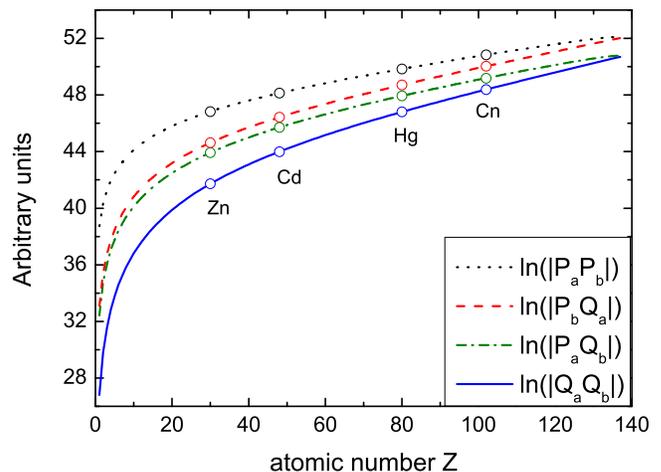}
\caption{(Color online): $Z$-dependence of the atomic EDM.
           The right hand side of the equation~(\ref{eq:ffgg2gamma}),
          calculated from (absolute values of)
	  one-electron wave function factors 
	 $P_a P_b $, $ Q_a Q_b$,
	 $P_a Q_b $, and $ P_b Q_a$.
	 The factors were generated from the
	 equations~(\ref{eq:ff}) and~(\ref{eq:gg}),
	 and evaluated at $ r = r_N $, as functions of atomic number $Z$.
%
 See text for details.
 }
\label{f1g1}
\end{figure}
%
{Neglecting a weak $Z$-dependence through the 
gamma function $2/{\Gamma(2 \gamma + 1)}$,
for small values of $Z$ the polynomial factor $Z^k$ determines
the functional form of the dependence on $Z$
while for large values of $Z$
the exponential factor $(2Zr_N)^{2 \gamma -2}$ takes over.
It can be shown numerically, 
as can bee seen in the Figure~\ref{f1g1},
that the polynomial $Z^k$ shape dominates up to about $Z=60$,
then in the range $ 60 < Z < 120 $ the function $ d_{at}(Z)$
is approximately exponential, and eventually the
approximation breaks down, because
the analytic approximation in
equations~\ref{eq:ff} and~\ref{eq:gg}
is valid only within the atomic number $Z$ range, where
bound solutions of the Dirac equation exist 
($Z \leq 137$ for point-like nuclei).}

%
%
{The analysis
above has been made {under} the assumption of a point-like
Coulomb field in the Dirac radial equation.
The finite sizes of nuclei entered only when the
equation~(\ref{eq:ffgg2gamma}) was evaluated.
For extended nuclei the solution of the Dirac equation} 
depends on the specific form
of the nuclear charge
distribution~\cite{GrantBook2007,GreinerBookRQM,GreinerBookQE}.
Flambaum and Ginges~\cite{FlambaumGinges:2002},
and Dzuba~et~al~\cite{Dzuba-Flambaum-Harabati-ratios:2011}
assumed uniform distribution of the electric charge inside a sphere
(with the normalization factors from~\cite{Khriplovich:1991}),
and obtained enhancement factors of a similar form as in
equations~(\ref{eq:K2gamma}) and~(\ref{eq:ffgg2gamma}), 
for angular symmetries
$s_{1/2}$, $p_{1/2}$, and $p_{3/2}$.

In the present paper
the numerical calculations
for  the {homologous} series of the group 12 of the periodic table
($^{69}_{30}$Zn,
 $^{111}_{\phantom{1}48}$Cd,
 $^{199}_{\phantom{1}80}$Hg,
 $^{285}_{112}$Cn, and 
 $^{482}_{162}$Uhb)
were performed with
extended nuclear model~(\ref{eq:fermi2}),
for which bound solutions of the Dirac-Fock equations
exist up to $Z=173$~\cite{Indelicato:Z-173:2011}.
} 
%
%
The dependence of EDMs on atomic number {$Z$}
along          group 12 of the periodic table
is presented in
Fig.~\ref{linZ_lnEDM}.
%
The red {circles} represent our final results,
calculated within the MCDHF-VOS.5 electron correlation model described above.
The blue {pluses} represent the uncorrelated DF results. 
The green plus in the upper right corner represents the EDM value
for Uhb.
Due to very large spin-orbit splitting of the $8p$ shell,
(see section~\ref{section.Copernicium} above),
the Uhb energy denominators are distinctively different from those of
other {members of the homologous series}.
To compensate for this splitting, we also
computed the EDMs for Uhb with energy denominators {taken} from Cn.
The latter value is represented by the square in 
Fig.~\ref{linZ_lnEDM}.
The solid line is fitted to the four (Zn, Cd, Hg, Cn) final results.
The dashed line  is fitted to the four uncorrelated DF results.
The Uhb results were excluded from the fitting.
{The regression {analysis}}
yields the following relations:
\begin{eqnarray} 
\label{eq:4fits} 
 \begin{aligned}
%
%
d^{TPT} \,~~~~~= [ -1.22(8) \cdot&e^{0.0766(6) \cdot Z}~~~~- 5(6)]
                                               \cdot 10^{-22} &\\ 
d^{PSS} \,~~~~~= [ ~-30(1) \cdot&e^{0.0813(3)  \cdot Z} - 8.54(1)]
                                               \cdot 10^{-26} &\\ 
d^{NSM} \,~~~~= [ -1.77(7) \cdot&e^{0.0626(3)  \cdot Z}~~~~+ 2(2)]
                                               \cdot 10^{-19} &\\
d^{eEDM}/\mu = [ ~~2.74(8) \cdot&e^{0.0841(2)  \cdot Z} ~~\,- 15(9)]
                                               \cdot 10^{-6~} &\\
 \end{aligned}
\end{eqnarray}
%
where the numbers in parentheses represent RMSE deviations.
The third line of the equation~(\ref{eq:4fits}) is displayed in
Fig.~\ref{linZ_lnEDM}.

{ Similar
regression
analysis can be done for the semi-analytic relations presented
for the point-nucleus case
in the equation~(\ref{eq:ffgg2gamma}) and in the Fig.~\ref{f1g1},
but restricted to the range of atomic numbers  $ 30 \leq Z \leq 112 $,
covered by the four elements (Zn, Cd, Hg, Cn) considered in this paper.
The analysis yields $d^{PSS} \sim e^{0.017 \cdot Z}$
                and $d^{NSM} \sim e^{0.022 \cdot Z}$,
somewhat smaller exponents than those presented in the
equation~(\ref{eq:4fits}).
} 
%
%
%

The deviation of the EDM value for the element E162
from the fitted function
in Figure~\ref{linZ_lnEDM} 
may be explained by several possible mechanisms:
rearrangements of the valence shells, i.e.~relativistic
contraction of the $8s$ and $8p_{1/2}$ shells, which results in the
above mentioned large spin-orbit splitting of the $8p$ shell;
variation of transition energy denominators, induced by
shell rearrangements;
contribution of QED effects, which could be quite
sizeable near the end of the periodic table
at {$Z=173$}~\cite{Goidenko:QED-112:2007,Indelicato:Z-173:2011}.
{However, the most likely explanation is the breakdown of the
exponential approximation near the end of the periodic table.
The analytic approximation in
equations~\ref{eq:ff} and~\ref{eq:gg}
is valid only within the atomic number $Z$ range, where
bound solutions of the Dirac equation exist 
( $Z \leq 137$ for point-like nuclei, $Z \leq 173$ for extended nuclei).
The element E162 is close to the end of the periodic table
at {$Z=173$}, where determination of a numerical {wave function},
even at the Dirac-Fock level,
may be problematic or impossible,
and one might expect a question
whether perturbation theory still works in
QED for elements close to Z = 173~\cite{Indelicato:Z-173:2011}.}

%
At very short distances  {$Z$}-dependence algebra is dominated by
the cutoff radii {$r_N$} (related to the sizes of the nuclei), and by
the power series solutions for $P_{n \kappa}$ and $Q_{n \kappa}$ at
the origin~\cite{GrantBook2007}.
The power series coefficients for $P_{n \kappa}$ and $Q_{n \kappa}$
depend on {the} nuclear potential
(again related to the sizes of the nuclei), and are constrained
by orthogonality of {the} one-electron functions {with the same symmetry}.
The dominant contributions to the 
matrix elements of the $ \hat{H}_{int} $ operators
come from the valence $ns^2$ orbitals in the ground state, 
and from the lowest $np_{1/2}$ and $np_{3/2}$ orbitals in the excited states.

%
\begin{figure}
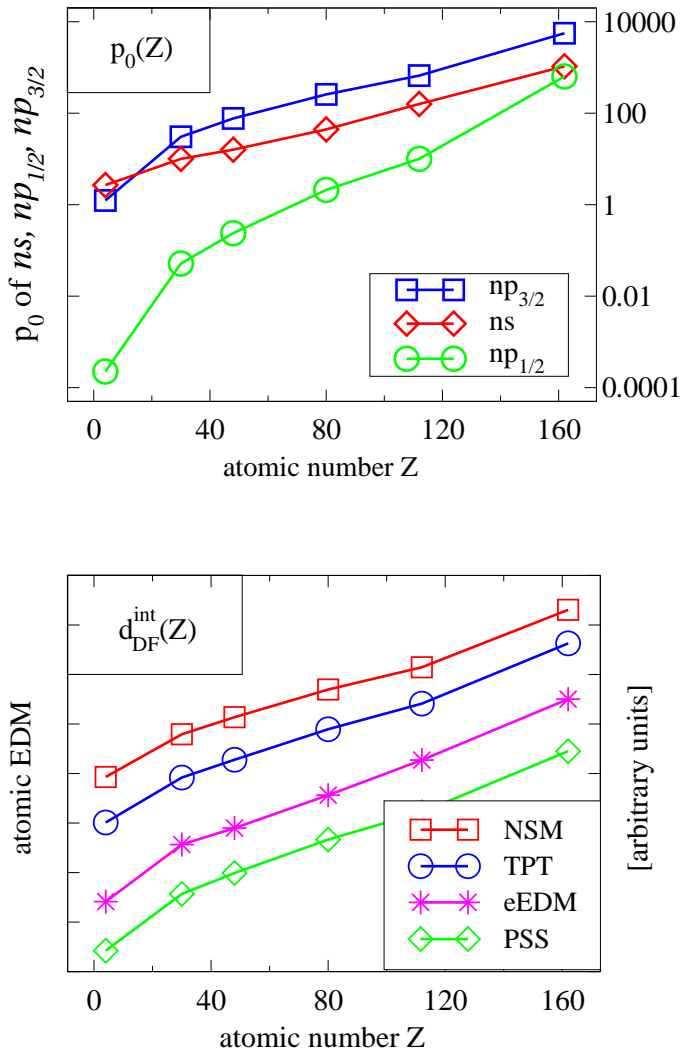

{
 \includegraphics[width=0.50\textwidth]{figure3a.eps}
 \vspace{44pt}  \phantom{ppp}
 \includegraphics[trim= 0cm 0cm 0cm 2cm,width=0.48\textwidth]{figure3b.eps}
}
\caption{(Color online)
Upper: power series coefficients $p_0$ of valence orbitals
 as functions of atomic number {$Z$}.
 Blue squares = $np_{3/2}$;
 red diamonds = $ns$;
 green circles = $np_{1/2}$;
 $n$ =  2, 4, 5, 6, 7, 8  for  Be, Zn, Cd, Hg, Cn, Uhb,  respectively.
Lower: atomic EDM (in arbitrary units on logarithmic scale) induced by
NSM (red squares),
TPT (blue circles),
eEDM (magenta stars),
and by
PSS (green diamonds),
as a function of atomic number {$Z$}.
All lines in both graphs are drawn only for the guidance of the eyes.
 See text for details.
}
\label{p0+d_DF} 
\end{figure}

The upper graph in the
Fig.~\ref{p0+d_DF} shows the coefficient $p_0$ of the lowest order
polynomial in the series expansion at the origin of the large component
$P$ of the radial function of the valence orbitals
($ns$, $np_{1/2}$, $np_{3/2}$)
of the elements from the group 12
(plus beryllium).
The quantum number $n$ assumes the values
2, 4, 5, 6, 7, 8 for Be, Zn, Cd, Hg, Cn, Uhb, respectively.
The lower graph in the
Fig.~\ref{p0+d_DF} shows the atomic EDMs induced by the 
TPT, PSS, NSM, and eEDM mechanisms,
as functions of atomic number {$Z$} for the elements of the group 12
(plus Be).
For the purpose of this comparison,
all values in Fig.~\ref{p0+d_DF} were obtained in the Dirac-Fock
approximation, without account of electron correlation effects,
and with the extended nuclear model~(\ref{eq:fermi2}).
Analogously to the point nucleus case
(represented by the equation~(\ref{eq:ffgg2gamma})),
the function $ d_{at}(Z)$ is approximately exponential
in the range $ 60 < Z < 120 $,
i.e.~for heavy and superheavy elements relevant from 
the point of view of the EDM searches.
%
%
Both graphs in
Fig.~\ref{p0+d_DF} show similar $Z$-dependencies
as those in
Figure~\ref{f1g1}, i.~.e.~the
polynomial shape up to about $Z=60$,
then approximately exponential
in the range $ 60 < Z < 120 $,
and eventually the exponential
approximation breaks down near
{the end of the extended
Periodic Table of Elements~\cite{Pyykko:Z-172:2011,Indelicato:Z-173:2011}
at $Z=173$.}
When comparing the shapes of the curves in the upper and lower graphs,
one has to bear in mind that radial integrals
in matrix elements of the $ \hat{H}_{int} $ operators
involve valence $ns^2$ orbitals in the ground state, 
and $np_{1/2}$ and $np_{3/2}$ orbitals in the excited states.
The apparent similarity of the $np_{1/2}$ and $np_{3/2}$ curves in the
upper graph and the four curves in the lower graph
is a numerical confirmation of the dominant role of power series
coefficients in the matrix element of the $ \hat{H}_{int} $ operators,
as well as 
of the proportionality relations between matrix elements, established
in~\cite{Dzuba-Flambaum-Harabati-ratios:2011}.
%
{Beryllium does not belong to the group 12
(which results in the visible deviation of Be from the fitted function)
but was included 
in Fig.~\ref{p0+d_DF} to indicate that the dominant role of power series
coefficients, as well as the proportionality
relations~\cite{Dzuba-Flambaum-Harabati-ratios:2011},
are not limited to one group of elements.}
{ The other deviations from linearity
 in the Figure~\ref{p0+d_DF}
 are induced by electron correlation effects,
 whose contributions  differ from element to element due to shell 
 rearrangements.}
%

\section{Conclusions}
\label{section.Conclusions}
%
{Electric dipole {moments (EDMs) have} not yet been detected experimentally.
The experimental searches {have been going on for} the last 50 years,
and the role of theory is not only to provide the limits on the fundamental
parameters, but also to guide the experimentalists to atoms, molecules,
and other systems with suitable enhancement factors.
Experimentalists need
to know (order of magnitude of) enhancement factors before they set up
their apparatus to detect EDM in a new species~\cite{Jungmann:2013}.
The present paper is intended to present the calculations
of EDMs, carried out with the multiconfiguration Dirac-Hartree-Fock theory,
of a superheavy element 
$^{285}_{112}$Cn.}
The main conclusion of the present paper is the suggestion for setting up
an EDM experiment on a superheavy element,
which would result in an order of magnitude increase of sensitivity, 
compared to a homologous heavy element.
Such {homologous} pairs include (but are not limited to)
%
%
\mbox{Yb--No}, 
\mbox{Hg--Cn}, 
\mbox{Tl--E113}, 
\mbox{Po--Lv}, 
\mbox{At--E117}, 
\mbox{Rn--E118}, 
\mbox{Fr--E119}, 
\mbox{Ra--E120}. 
%
%
%
If the exponential {$Z$}-dependence derived in the present paper is assumed
for all above {homologous} pairs, 
an increase of sensitivity by a factor 8-30 should be expected.
The best limit on the EDM of a diamagnetic atom
comes from $^{199}$Hg, for which 
\mbox{ $ d(^{199}{\mbox{Hg}}) < 3.1 \times 10^{-29} e\cdot$cm} (95\%~C.L.)
has been reported~\cite{edm-199-Hg-Griffith-2009}.
Our calculations indicate that for the \mbox{Hg--Cn} pair
the increase of sensitivity would be
57.5/4.8,
236.9/17.0,
17.6/2.2,
and 
299.7/10.0
for TPT, PSS, NSM, and eEDM, respectively.
%
%
%
Over the last 50 years the precision of EDM experiments
has been improving by about an order of magnitude per
decade~\cite{Gould:2007,Jungmann-2013,%
edm-199-Hg-Romalis-2001,edm-199-Hg-Griffith-2009,edm-199-Hg-Swallows:2013,%
Baron:2014}.
On this timescale an experiment on Cn would be
equivalent to
{time travel into the future over a distance of about ten to twenty years.} 

{We are of course aware of the fact that an EDM experiment on a
short-lived superheavy element is impractical at this time.
However, the {techniques for}}
trapping atoms~\cite{Wieman:trapping:1999,Willmann:barium:2015},
controlling quantum
systems~\cite{Wineland:nobellecture:2013,Ludlow:clocks:2015},
{and performing spectroscopic investigations of
radioactive~\cite{Lu:edm-Ra:2015}
and superheavy elements~\cite{Cheal:progress:2010}}
{advance rapidly. {At the same time}
the quest for the superheavy island of stability
continues~\cite{Oganessian:island:2012,Zagrebaev:2013},
and sooner or later one may expect a breakthrough of laser {spectroscopic}
methods into the realm of
superheavy elements~\cite{Cheal:progress:2010}.}

%
%
{The EDM experiments with superheavy elements, if ever becoming
feasible,
would probably constitute the final frontier for atomic tests of
violation of parity ($P$) and time reversal ($T$)
symmetries.
In recents years the molecular avenue promises to become more competitive
in the EDM searches.
The advantages of molecular eEDM experiments is in
the large values of the effective electric field,
several orders of magnitude higher than those in
atoms~\cite{Hinds:2011,Baron:2014,Sasmal:2016}.
%
Current progress in cooling and trapping
molecules~\cite{Norrgard-DeMille:2016,Prehn:2016,Isaev-Berger:2016,%
Panda-Gabrielse:2016,Fleig:2016},
%
%
as well as
molecular ions~\cite{Leanhardt-Cornell:2011,Gresh-Cornell:2016},
%
may soon allow to increase coherence times and improve population control
in molecular experiments,
%
which might translate into a significant advantage of molecular experiments 
over atomic ones.
} 

{While an EDM experiment on a
short-lived superheavy element is impractical at this time,
still less practical would be an experiment on a superheavy molecule.
However, when molecular EDM experiments become feasible,
one may also envisage making, trapping, 
and eventually performing spectroscopy of superheavy molecules.
%
It is difficult to say what is impossible,
for the dream of yesterday is the hope of today and the reality of
tomorrow~\cite{Goddard}.
} 

\begin{acknowledgments}
%
LR thanks for the HPC resources provided by the ITOAC of Vilnius University.
%
JB,~GG,~PJ~acknowledge the support from
the Polish Ministry of Science and Higher Education (MNiSW)
in the framework of the grant No.~N~N202~014140
awarded for the years 2011-2016.
JB~acknowledges the support from
the European Regional
Development Fund in the framework of the Polish Innovation
Economy Operational Program (contract no. POIG.02.01.00-12-023/08).
{PJ~acknowledges financial support from the Swedish Research Council
under the Grant 2015-04842.} 
\end{acknowledgments}


\bibliography{xet}

\begin{thebibliography}{97}%
\makeatletter
\providecommand \@ifxundefined [1]{%
 \@ifx{#1\undefined}
}%
\providecommand \@ifnum [1]{%
 \ifnum #1\expandafter \@firstoftwo
 \else \expandafter \@secondoftwo
 \fi
}%
\providecommand \@ifx [1]{%
 \ifx #1\expandafter \@firstoftwo
 \else \expandafter \@secondoftwo
 \fi
}%
\providecommand \natexlab [1]{#1}%
\providecommand \enquote  [1]{``#1''}%
\providecommand \bibnamefont  [1]{#1}%
\providecommand \bibfnamefont [1]{#1}%
\providecommand \citenamefont [1]{#1}%
\providecommand \href@noop [0]{\@secondoftwo}%
\providecommand \href [0]{\begingroup \@sanitize@url \@href}%
\providecommand \@href[1]{\@@startlink{#1}\@@href}%
\providecommand \@@href[1]{\endgroup#1\@@endlink}%
\providecommand \@sanitize@url [0]{\catcode `\\12\catcode `\$12\catcode
  `\&12\catcode `\#12\catcode `\^12\catcode `\_12\catcode `\%12\relax}%
\providecommand \@@startlink[1]{}%
\providecommand \@@endlink[0]{}%
\providecommand \url  [0]{\begingroup\@sanitize@url \@url }%
\providecommand \@url [1]{\endgroup\@href {#1}{\urlprefix }}%
\providecommand \urlprefix  [0]{URL }%
\providecommand \Eprint [0]{\href }%
\providecommand \doibase [0]{http://dx.doi.org/}%
\providecommand \selectlanguage [0]{\@gobble}%
\providecommand \bibinfo  [0]{\@secondoftwo}%
\providecommand \bibfield  [0]{\@secondoftwo}%
\providecommand \translation [1]{[#1]}%
\providecommand \BibitemOpen [0]{}%
\providecommand \bibitemStop [0]{}%
\providecommand \bibitemNoStop [0]{.\EOS\space}%
\providecommand \EOS [0]{\spacefactor3000\relax}%
\providecommand \BibitemShut  [1]{\csname bibitem#1\endcsname}%
\let\auto@bib@innerbib\@empty
\bibitem [{\citenamefont {Khriplovich}\ and\ \citenamefont
  {Lamoreaux}(1997)}]{KhriplovichLamoreaux:1997}%
  \BibitemOpen
  \bibfield  {author} {\bibinfo {author} {\bibfnamefont {I.~B.}\ \bibnamefont
  {Khriplovich}}\ and\ \bibinfo {author} {\bibfnamefont {S.~K.}\ \bibnamefont
  {Lamoreaux}},\ }\href@noop {} {\emph {\bibinfo {title} {CP Violation Without
  Strangeness}}}\ (\bibinfo  {publisher} {Springer},\ \bibinfo {address}
  {Berlin},\ \bibinfo {year} {1997})\BibitemShut {NoStop}%
\bibitem [{\citenamefont {Jungmann}(2013{\natexlab{a}})}]{Jungmann-2013}%
  \BibitemOpen
  \bibfield  {author} {\bibinfo {author} {\bibfnamefont {K.}~\bibnamefont
  {Jungmann}},\ }\href@noop {} {\bibfield  {journal} {\bibinfo  {journal}
  {Ann.~Phys.}\ }\textbf {\bibinfo {volume} {525}},\ \bibinfo {pages}
  {550–564} (\bibinfo {year} {2013}{\natexlab{a}})}\BibitemShut {NoStop}%
\bibitem [{\citenamefont {Wu}\ \emph {et~al.}(1957)\citenamefont {Wu},
  \citenamefont {Ambler}, \citenamefont {Hayward}, \citenamefont {Hoppes},\
  and\ \citenamefont {Hudson}}]{Wu:1957}%
  \BibitemOpen
  \bibfield  {author} {\bibinfo {author} {\bibfnamefont {C.~S.}\ \bibnamefont
  {Wu}}, \bibinfo {author} {\bibfnamefont {E.}~\bibnamefont {Ambler}}, \bibinfo
  {author} {\bibfnamefont {R.~W.}\ \bibnamefont {Hayward}}, \bibinfo {author}
  {\bibfnamefont {D.~D.}\ \bibnamefont {Hoppes}}, \ and\ \bibinfo {author}
  {\bibfnamefont {R.~P.}\ \bibnamefont {Hudson}},\ }\href@noop {} {\bibfield
  {journal} {\bibinfo  {journal} {Phys.~Rev.}\ }\textbf {\bibinfo {volume}
  {105}},\ \bibinfo {pages} {1413} (\bibinfo {year} {1957})}\BibitemShut
  {NoStop}%
\bibitem [{\citenamefont {Garwin}\ \emph {et~al.}(1957)\citenamefont {Garwin},
  \citenamefont {Lederman},\ and\ \citenamefont {Weinrich}}]{Garwin:muon:1957}%
  \BibitemOpen
  \bibfield  {author} {\bibinfo {author} {\bibfnamefont {R.~L.}\ \bibnamefont
  {Garwin}}, \bibinfo {author} {\bibfnamefont {L.~M.}\ \bibnamefont
  {Lederman}}, \ and\ \bibinfo {author} {\bibfnamefont {M.}~\bibnamefont
  {Weinrich}},\ }\href@noop {} {\bibfield  {journal} {\bibinfo  {journal}
  {Phys.~Rev.}\ }\textbf {\bibinfo {volume} {105}},\ \bibinfo {pages} {1415}
  (\bibinfo {year} {1957})}\BibitemShut {NoStop}%
\bibitem [{\citenamefont {Friedman}\ and\ \citenamefont
  {Telegdi}(1957)}]{Friedman:pion:1957}%
  \BibitemOpen
  \bibfield  {author} {\bibinfo {author} {\bibfnamefont {J.~I.}\ \bibnamefont
  {Friedman}}\ and\ \bibinfo {author} {\bibfnamefont {V.~L.}\ \bibnamefont
  {Telegdi}},\ }\href@noop {} {\bibfield  {journal} {\bibinfo  {journal}
  {Phys.~Rev.}\ }\textbf {\bibinfo {volume} {106}},\ \bibinfo {pages} {1290}
  (\bibinfo {year} {1957})}\BibitemShut {NoStop}%
\bibitem [{\citenamefont {Christenson}\ \emph {et~al.}(1964)\citenamefont
  {Christenson}, \citenamefont {Cronin}, \citenamefont {Fitch},\ and\
  \citenamefont {Turlay}}]{Christenson:kaon:1957}%
  \BibitemOpen
  \bibfield  {author} {\bibinfo {author} {\bibfnamefont {J.~H.}\ \bibnamefont
  {Christenson}}, \bibinfo {author} {\bibfnamefont {J.~W.}\ \bibnamefont
  {Cronin}}, \bibinfo {author} {\bibfnamefont {V.~L.}\ \bibnamefont {Fitch}}, \
  and\ \bibinfo {author} {\bibfnamefont {R.}~\bibnamefont {Turlay}},\
  }\href@noop {} {\bibfield  {journal} {\bibinfo  {journal} {Phys.~Rev.~Lett.}\
  }\textbf {\bibinfo {volume} {13}},\ \bibinfo {pages} {138} (\bibinfo {year}
  {1964})}\BibitemShut {NoStop}%
\bibitem [{\citenamefont {Kostelecky}\ and\ \citenamefont
  {Russell}(2011)}]{Kostelecky-2011}%
  \BibitemOpen
  \bibfield  {author} {\bibinfo {author} {\bibfnamefont {V.~A.}\ \bibnamefont
  {Kostelecky}}\ and\ \bibinfo {author} {\bibfnamefont {N.}~\bibnamefont
  {Russell}},\ }\href@noop {} {\bibfield  {journal} {\bibinfo  {journal}
  {Rev.~Mod.~Phys.}\ }\textbf {\bibinfo {volume} {83}},\ \bibinfo {pages} {1}
  (\bibinfo {year} {2011})}\BibitemShut {NoStop}%
\bibitem [{\citenamefont {Angelopoulos}\ \emph {et~al.}(1998)\citenamefont
  {Angelopoulos}, \citenamefont {Apostolakis}, \citenamefont {Aslanides},
  \citenamefont {Backenstoss}, \citenamefont {Bargassa}, \citenamefont
  {Behnke}, \citenamefont {Benelli}, \citenamefont {Bertin}, \citenamefont
  {Blanc}, \citenamefont {Bloch}, \citenamefont {Carlson}, \citenamefont
  {Carroll}, \citenamefont {Cawley}, \citenamefont {Charalambous},
  \citenamefont {Chertok}, \citenamefont {Danielsson}, \citenamefont
  {Dejardin}, \citenamefont {Derre}, \citenamefont {Ealet}, \citenamefont
  {Eleftheriadis}, \citenamefont {Faravel}, \citenamefont {Fetscher},
  \citenamefont {Fidecaro}, \citenamefont {Filip\u{c}i\u{c}}, \citenamefont
  {Francis}, \citenamefont {Fry}, \citenamefont {Gabathuler}, \citenamefont
  {Gamet}, \citenamefont {Gerber}, \citenamefont {Go}, \citenamefont
  {Haselden}, \citenamefont {Hayman}, \citenamefont {Henry-Couannier},
  \citenamefont {Hollander}, \citenamefont {Jon-And}, \citenamefont {Kettle},
  \citenamefont {Kokkas}, \citenamefont {Kreuger}, \citenamefont {Gac},
  \citenamefont {Leimgruber}, \citenamefont {Mandi\'{c}}, \citenamefont
  {Manthos}, \citenamefont {Marel}, \citenamefont {Miku\u{z}}, \citenamefont
  {Miller}, \citenamefont {Montanet}, \citenamefont {Muller}, \citenamefont
  {Nakada}, \citenamefont {Pagels}, \citenamefont {Papadopoulos}, \citenamefont
  {Pavlopoulos}, \citenamefont {Policarpo}, \citenamefont {Polivka},
  \citenamefont {Rickenbach}, \citenamefont {Roberts}, \citenamefont {Ruf},
  \citenamefont {Santoni}, \citenamefont {Sch{\aa}fer}, \citenamefont
  {Schaller}, \citenamefont {Schietinger}, \citenamefont {Schopper},
  \citenamefont {Tauscher}, \citenamefont {Thibault}, \citenamefont {Touchard},
  \citenamefont {Touramanis}, \citenamefont {Eijk}, \citenamefont {Vlachos},
  \citenamefont {Weber}, \citenamefont {Wigger}, \citenamefont {Wolter},
  \citenamefont {Zavrtanik},\ and\ \citenamefont {Zimmerman}}]{CPLEAR:1998}%
  \BibitemOpen
  \bibfield  {author} {\bibinfo {author} {\bibfnamefont {A.}~\bibnamefont
  {Angelopoulos}}, \bibinfo {author} {\bibfnamefont {A.}~\bibnamefont
  {Apostolakis}}, \bibinfo {author} {\bibfnamefont {E.}~\bibnamefont
  {Aslanides}}, \bibinfo {author} {\bibfnamefont {G.}~\bibnamefont
  {Backenstoss}}, \bibinfo {author} {\bibfnamefont {P.}~\bibnamefont
  {Bargassa}}, \bibinfo {author} {\bibfnamefont {O.}~\bibnamefont {Behnke}},
  \bibinfo {author} {\bibfnamefont {A.}~\bibnamefont {Benelli}}, \bibinfo
  {author} {\bibfnamefont {V.}~\bibnamefont {Bertin}}, \bibinfo {author}
  {\bibfnamefont {F.}~\bibnamefont {Blanc}}, \bibinfo {author} {\bibfnamefont
  {P.}~\bibnamefont {Bloch}}, \bibinfo {author} {\bibfnamefont
  {P.}~\bibnamefont {Carlson}}, \bibinfo {author} {\bibfnamefont
  {M.}~\bibnamefont {Carroll}}, \bibinfo {author} {\bibfnamefont
  {E.}~\bibnamefont {Cawley}}, \bibinfo {author} {\bibfnamefont
  {S.}~\bibnamefont {Charalambous}}, \bibinfo {author} {\bibfnamefont
  {M.}~\bibnamefont {Chertok}}, \bibinfo {author} {\bibfnamefont
  {M.}~\bibnamefont {Danielsson}}, \bibinfo {author} {\bibfnamefont
  {M.}~\bibnamefont {Dejardin}}, \bibinfo {author} {\bibfnamefont
  {J.}~\bibnamefont {Derre}}, \bibinfo {author} {\bibfnamefont
  {A.}~\bibnamefont {Ealet}}, \bibinfo {author} {\bibfnamefont
  {C.}~\bibnamefont {Eleftheriadis}}, \bibinfo {author} {\bibfnamefont
  {L.}~\bibnamefont {Faravel}}, \bibinfo {author} {\bibfnamefont
  {W.}~\bibnamefont {Fetscher}}, \bibinfo {author} {\bibfnamefont
  {M.}~\bibnamefont {Fidecaro}}, \bibinfo {author} {\bibfnamefont
  {A.}~\bibnamefont {Filip\u{c}i\u{c}}}, \bibinfo {author} {\bibfnamefont
  {D.}~\bibnamefont {Francis}}, \bibinfo {author} {\bibfnamefont
  {J.}~\bibnamefont {Fry}}, \bibinfo {author} {\bibfnamefont {E.}~\bibnamefont
  {Gabathuler}}, \bibinfo {author} {\bibfnamefont {R.}~\bibnamefont {Gamet}},
  \bibinfo {author} {\bibfnamefont {H.-J.}\ \bibnamefont {Gerber}}, \bibinfo
  {author} {\bibfnamefont {A.}~\bibnamefont {Go}}, \bibinfo {author}
  {\bibfnamefont {A.}~\bibnamefont {Haselden}}, \bibinfo {author}
  {\bibfnamefont {P.}~\bibnamefont {Hayman}}, \bibinfo {author} {\bibfnamefont
  {F.}~\bibnamefont {Henry-Couannier}}, \bibinfo {author} {\bibfnamefont
  {R.}~\bibnamefont {Hollander}}, \bibinfo {author} {\bibfnamefont
  {K.}~\bibnamefont {Jon-And}}, \bibinfo {author} {\bibfnamefont {P.-R.}\
  \bibnamefont {Kettle}}, \bibinfo {author} {\bibfnamefont {P.}~\bibnamefont
  {Kokkas}}, \bibinfo {author} {\bibfnamefont {R.}~\bibnamefont {Kreuger}},
  \bibinfo {author} {\bibfnamefont {R.~L.}\ \bibnamefont {Gac}}, \bibinfo
  {author} {\bibfnamefont {F.}~\bibnamefont {Leimgruber}}, \bibinfo {author}
  {\bibfnamefont {I.}~\bibnamefont {Mandi\'{c}}}, \bibinfo {author}
  {\bibfnamefont {N.}~\bibnamefont {Manthos}}, \bibinfo {author} {\bibfnamefont
  {G.}~\bibnamefont {Marel}}, \bibinfo {author} {\bibfnamefont
  {M.}~\bibnamefont {Miku\u{z}}}, \bibinfo {author} {\bibfnamefont
  {J.}~\bibnamefont {Miller}}, \bibinfo {author} {\bibfnamefont
  {F.}~\bibnamefont {Montanet}}, \bibinfo {author} {\bibfnamefont
  {A.}~\bibnamefont {Muller}}, \bibinfo {author} {\bibfnamefont
  {T.}~\bibnamefont {Nakada}}, \bibinfo {author} {\bibfnamefont
  {B.}~\bibnamefont {Pagels}}, \bibinfo {author} {\bibfnamefont
  {I.}~\bibnamefont {Papadopoulos}}, \bibinfo {author} {\bibfnamefont
  {P.}~\bibnamefont {Pavlopoulos}}, \bibinfo {author} {\bibfnamefont
  {A.}~\bibnamefont {Policarpo}}, \bibinfo {author} {\bibfnamefont
  {G.}~\bibnamefont {Polivka}}, \bibinfo {author} {\bibfnamefont
  {R.}~\bibnamefont {Rickenbach}}, \bibinfo {author} {\bibfnamefont
  {B.}~\bibnamefont {Roberts}}, \bibinfo {author} {\bibfnamefont
  {T.}~\bibnamefont {Ruf}}, \bibinfo {author} {\bibfnamefont {C.}~\bibnamefont
  {Santoni}}, \bibinfo {author} {\bibfnamefont {M.}~\bibnamefont
  {Sch{\aa}fer}}, \bibinfo {author} {\bibfnamefont {L.}~\bibnamefont
  {Schaller}}, \bibinfo {author} {\bibfnamefont {T.}~\bibnamefont
  {Schietinger}}, \bibinfo {author} {\bibfnamefont {A.}~\bibnamefont
  {Schopper}}, \bibinfo {author} {\bibfnamefont {L.}~\bibnamefont {Tauscher}},
  \bibinfo {author} {\bibfnamefont {C.}~\bibnamefont {Thibault}}, \bibinfo
  {author} {\bibfnamefont {F.}~\bibnamefont {Touchard}}, \bibinfo {author}
  {\bibfnamefont {C.}~\bibnamefont {Touramanis}}, \bibinfo {author}
  {\bibfnamefont {C.~V.}\ \bibnamefont {Eijk}}, \bibinfo {author}
  {\bibfnamefont {S.}~\bibnamefont {Vlachos}}, \bibinfo {author} {\bibfnamefont
  {P.}~\bibnamefont {Weber}}, \bibinfo {author} {\bibfnamefont
  {O.}~\bibnamefont {Wigger}}, \bibinfo {author} {\bibfnamefont
  {M.}~\bibnamefont {Wolter}}, \bibinfo {author} {\bibfnamefont
  {D.}~\bibnamefont {Zavrtanik}}, \ and\ \bibinfo {author} {\bibfnamefont
  {D.}~\bibnamefont {Zimmerman}} (\bibinfo {collaboration} {CPLEAR
  Collaboration}),\ }\href@noop {} {\bibfield  {journal} {\bibinfo  {journal}
  {Phys.~Lett.~B}\ }\textbf {\bibinfo {volume} {444}},\ \bibinfo {pages} {43}
  (\bibinfo {year} {1998})}\BibitemShut {NoStop}%
\bibitem [{\citenamefont {Wolfenstein}(1999)}]{Wolfenstein:1999}%
  \BibitemOpen
  \bibfield  {author} {\bibinfo {author} {\bibfnamefont {L.}~\bibnamefont
  {Wolfenstein}},\ }\href@noop {} {\bibfield  {journal} {\bibinfo  {journal}
  {Int.~J.~Mod.~Phys.~E}\ }\textbf {\bibinfo {volume} {08}},\ \bibinfo {pages}
  {501} (\bibinfo {year} {1999})}\BibitemShut {NoStop}%
\bibitem [{\citenamefont {Bernabeu}\ \emph {et~al.}(2013)\citenamefont
  {Bernabeu}, \citenamefont {Domenico},\ and\ \citenamefont
  {Villanueva-Perez}}]{Bernabeu:2013}%
  \BibitemOpen
  \bibfield  {author} {\bibinfo {author} {\bibfnamefont {J.}~\bibnamefont
  {Bernabeu}}, \bibinfo {author} {\bibfnamefont {A.~D.}\ \bibnamefont
  {Domenico}}, \ and\ \bibinfo {author} {\bibfnamefont {P.}~\bibnamefont
  {Villanueva-Perez}},\ }\href@noop {} {\bibfield  {journal} {\bibinfo
  {journal} {Nucl.~Phys.~B}\ }\textbf {\bibinfo {volume} {868}},\ \bibinfo
  {pages} {102} (\bibinfo {year} {2013})}\BibitemShut {NoStop}%
\bibitem [{\citenamefont {{Lees \textit{et al.}}}(2012)}]{BABAR:2012}%
  \BibitemOpen
  \bibfield  {author} {\bibinfo {author} {\bibfnamefont {J.~P.}\ \bibnamefont
  {{Lees \textit{et al.}}}} (\bibinfo {collaboration} {The BABAR
  Collaboration}),\ }\href@noop {} {\bibfield  {journal} {\bibinfo  {journal}
  {Phys.~Rev.~Lett.}\ }\textbf {\bibinfo {volume} {109}},\ \bibinfo {pages}
  {211801} (\bibinfo {year} {2012})}\BibitemShut {NoStop}%
\bibitem [{\citenamefont {Cabibbo}(1963)}]{Cabibbo:1963}%
  \BibitemOpen
  \bibfield  {author} {\bibinfo {author} {\bibfnamefont {N.}~\bibnamefont
  {Cabibbo}},\ }\href@noop {} {\bibfield  {journal} {\bibinfo  {journal}
  {Phys.~Rev.~Lett.}\ }\textbf {\bibinfo {volume} {10}},\ \bibinfo {pages}
  {531} (\bibinfo {year} {1963})}\BibitemShut {NoStop}%
\bibitem [{\citenamefont {Sozzi}(2008)}]{Sozzi:book:2008}%
  \BibitemOpen
  \bibfield  {author} {\bibinfo {author} {\bibfnamefont {M.~S.}\ \bibnamefont
  {Sozzi}},\ }\href@noop {} {\emph {\bibinfo {title} {Discrete Symmetries and
  CP Violation. From Experiment to Theory}}}\ (\bibinfo  {publisher} {Oxford
  University Press},\ \bibinfo {address} {Oxford},\ \bibinfo {year}
  {2008})\BibitemShut {NoStop}%
\bibitem [{\citenamefont {Baker}\ \emph {et~al.}(2006)\citenamefont {Baker},
  \citenamefont {Doyle}, \citenamefont {Geltenbort}, \citenamefont {Green},
  \citenamefont {van~der Grinten}, \citenamefont {Harris}, \citenamefont
  {Iaydjiev}, \citenamefont {Ivanov}, \citenamefont {May}, \citenamefont
  {Pendlebury}, \citenamefont {Richardson}, \citenamefont {Shiers},\ and\
  \citenamefont {Smith}}]{Baker:2006}%
  \BibitemOpen
  \bibfield  {author} {\bibinfo {author} {\bibfnamefont {C.~A.}\ \bibnamefont
  {Baker}}, \bibinfo {author} {\bibfnamefont {D.~D.}\ \bibnamefont {Doyle}},
  \bibinfo {author} {\bibfnamefont {P.}~\bibnamefont {Geltenbort}}, \bibinfo
  {author} {\bibfnamefont {K.}~\bibnamefont {Green}}, \bibinfo {author}
  {\bibfnamefont {M.~G.~D.}\ \bibnamefont {van~der Grinten}}, \bibinfo {author}
  {\bibfnamefont {P.~G.}\ \bibnamefont {Harris}}, \bibinfo {author}
  {\bibfnamefont {P.}~\bibnamefont {Iaydjiev}}, \bibinfo {author}
  {\bibfnamefont {S.~N.}\ \bibnamefont {Ivanov}}, \bibinfo {author}
  {\bibfnamefont {D.~J.~R.}\ \bibnamefont {May}}, \bibinfo {author}
  {\bibfnamefont {J.~M.}\ \bibnamefont {Pendlebury}}, \bibinfo {author}
  {\bibfnamefont {J.~D.}\ \bibnamefont {Richardson}}, \bibinfo {author}
  {\bibfnamefont {D.}~\bibnamefont {Shiers}}, \ and\ \bibinfo {author}
  {\bibfnamefont {K.~F.}\ \bibnamefont {Smith}},\ }\href@noop {} {\bibfield
  {journal} {\bibinfo  {journal} {Phys.~Rev.~Lett.}\ }\textbf {\bibinfo
  {volume} {97}},\ \bibinfo {pages} {131801} (\bibinfo {year}
  {2006})}\BibitemShut {NoStop}%
\bibitem [{\citenamefont {Regan}\ \emph {et~al.}(2002)\citenamefont {Regan},
  \citenamefont {Commins}, \citenamefont {Schmidt},\ and\ \citenamefont
  {DeMille}}]{Regan:2002}%
  \BibitemOpen
  \bibfield  {author} {\bibinfo {author} {\bibfnamefont {B.~C.}\ \bibnamefont
  {Regan}}, \bibinfo {author} {\bibfnamefont {E.~D.}\ \bibnamefont {Commins}},
  \bibinfo {author} {\bibfnamefont {C.~J.}\ \bibnamefont {Schmidt}}, \ and\
  \bibinfo {author} {\bibfnamefont {D.}~\bibnamefont {DeMille}},\ }\href@noop
  {} {\bibfield  {journal} {\bibinfo  {journal} {Phys.~Rev.~Lett.}\ }\textbf
  {\bibinfo {volume} {88}},\ \bibinfo {pages} {071805} (\bibinfo {year}
  {2002})}\BibitemShut {NoStop}%
\bibitem [{\citenamefont {Griffith}\ \emph
  {et~al.}(2009{\natexlab{a}})\citenamefont {Griffith}, \citenamefont
  {Swallows}, \citenamefont {Loftus}, \citenamefont {Romalis}, \citenamefont
  {Heckel},\ and\ \citenamefont {Fortson}}]{Griffith:2009}%
  \BibitemOpen
  \bibfield  {author} {\bibinfo {author} {\bibfnamefont {W.~C.}\ \bibnamefont
  {Griffith}}, \bibinfo {author} {\bibfnamefont {M.~D.}\ \bibnamefont
  {Swallows}}, \bibinfo {author} {\bibfnamefont {T.~H.}\ \bibnamefont
  {Loftus}}, \bibinfo {author} {\bibfnamefont {M.~V.}\ \bibnamefont {Romalis}},
  \bibinfo {author} {\bibfnamefont {B.~R.}\ \bibnamefont {Heckel}}, \ and\
  \bibinfo {author} {\bibfnamefont {E.~N.}\ \bibnamefont {Fortson}},\
  }\href@noop {} {\bibfield  {journal} {\bibinfo  {journal} {Phys.~Rev.~Lett.}\
  }\textbf {\bibinfo {volume} {102}},\ \bibinfo {pages} {101601} (\bibinfo
  {year} {2009}{\natexlab{a}})}\BibitemShut {NoStop}%
\bibitem [{\citenamefont {Hudson}\ \emph {et~al.}(2011)\citenamefont {Hudson},
  \citenamefont {Kara}, \citenamefont {Smallman}, \citenamefont {Sauer},
  \citenamefont {Tarbutt},\ and\ \citenamefont {Hinds}}]{Hinds:2011}%
  \BibitemOpen
  \bibfield  {author} {\bibinfo {author} {\bibfnamefont {J.~J.}\ \bibnamefont
  {Hudson}}, \bibinfo {author} {\bibfnamefont {D.~M.}\ \bibnamefont {Kara}},
  \bibinfo {author} {\bibfnamefont {I.~J.}\ \bibnamefont {Smallman}}, \bibinfo
  {author} {\bibfnamefont {B.~E.}\ \bibnamefont {Sauer}}, \bibinfo {author}
  {\bibfnamefont {M.~R.}\ \bibnamefont {Tarbutt}}, \ and\ \bibinfo {author}
  {\bibfnamefont {E.~A.}\ \bibnamefont {Hinds}},\ }\href@noop {} {\bibfield
  {journal} {\bibinfo  {journal} {Nature}\ }\textbf {\bibinfo {volume} {473}},\
  \bibinfo {pages} {493} (\bibinfo {year} {2011})}\BibitemShut {NoStop}%
\bibitem [{\citenamefont {Baron}\ \emph {et~al.}(2014)\citenamefont {Baron},
  \citenamefont {Campbell}, \citenamefont {DeMille}, \citenamefont {Doyle},
  \citenamefont {Gabrielse}, \citenamefont {Gurevich}, \citenamefont {Hess},
  \citenamefont {Hutzler}, \citenamefont {Kirilov}, \citenamefont {Kozyryev},
  \citenamefont {O'Leary}, \citenamefont {Panda}, \citenamefont {Parsons},
  \citenamefont {Petrik}, \citenamefont {Spaun}, \citenamefont {Vutha},\ and\
  \citenamefont {West}}]{Baron:2014}%
  \BibitemOpen
  \bibfield  {author} {\bibinfo {author} {\bibfnamefont {T.~A. C.~J.}\
  \bibnamefont {Baron}}, \bibinfo {author} {\bibfnamefont {W.~C.}\ \bibnamefont
  {Campbell}}, \bibinfo {author} {\bibfnamefont {D.}~\bibnamefont {DeMille}},
  \bibinfo {author} {\bibfnamefont {J.~M.}\ \bibnamefont {Doyle}}, \bibinfo
  {author} {\bibfnamefont {G.}~\bibnamefont {Gabrielse}}, \bibinfo {author}
  {\bibfnamefont {Y.~V.}\ \bibnamefont {Gurevich}}, \bibinfo {author}
  {\bibfnamefont {P.~W.}\ \bibnamefont {Hess}}, \bibinfo {author}
  {\bibfnamefont {N.~R.}\ \bibnamefont {Hutzler}}, \bibinfo {author}
  {\bibfnamefont {E.}~\bibnamefont {Kirilov}}, \bibinfo {author} {\bibfnamefont
  {I.}~\bibnamefont {Kozyryev}}, \bibinfo {author} {\bibfnamefont {B.~R.}\
  \bibnamefont {O'Leary}}, \bibinfo {author} {\bibfnamefont {C.~D.}\
  \bibnamefont {Panda}}, \bibinfo {author} {\bibfnamefont {M.~F.}\ \bibnamefont
  {Parsons}}, \bibinfo {author} {\bibfnamefont {E.~S.}\ \bibnamefont {Petrik}},
  \bibinfo {author} {\bibfnamefont {B.}~\bibnamefont {Spaun}}, \bibinfo
  {author} {\bibfnamefont {A.~C.}\ \bibnamefont {Vutha}}, \ and\ \bibinfo
  {author} {\bibfnamefont {A.~D.}\ \bibnamefont {West}} (\bibinfo
  {collaboration} {The ACME Collaboration}),\ }\href@noop {} {\bibfield
  {journal} {\bibinfo  {journal} {Science}\ }\textbf {\bibinfo {volume}
  {343}},\ \bibinfo {pages} {269} (\bibinfo {year} {2014})}\BibitemShut
  {NoStop}%
\bibitem [{\citenamefont {Ginges}\ and\ \citenamefont
  {Flambaum}(2004)}]{Ginges:2004}%
  \BibitemOpen
  \bibfield  {author} {\bibinfo {author} {\bibfnamefont {J.~S.~M.}\
  \bibnamefont {Ginges}}\ and\ \bibinfo {author} {\bibfnamefont {V.~V.}\
  \bibnamefont {Flambaum}},\ }\href@noop {} {\bibfield  {journal} {\bibinfo
  {journal} {Phys. Rep.}\ }\textbf {\bibinfo {volume} {397}},\ \bibinfo {pages}
  {63} (\bibinfo {year} {2004})}\BibitemShut {NoStop}%
\bibitem [{\citenamefont {Roberts}\ and\ \citenamefont
  {Marciano}(2009)}]{RobertsMarciano:2009}%
  \BibitemOpen
  \bibinfo {editor} {\bibfnamefont {B.~L.}\ \bibnamefont {Roberts}}\ and\
  \bibinfo {editor} {\bibfnamefont {W.~J.}\ \bibnamefont {Marciano}},\ eds.,\
  \href@noop {} {\emph {\bibinfo {title} {Advanced Series on Directions in High
  Energy Physics}}},\ Vol.~\bibinfo {volume} {20}\ (\bibinfo  {publisher}
  {World Scientific},\ \bibinfo {address} {Singapore},\ \bibinfo {year}
  {2009})\BibitemShut {NoStop}%
\bibitem [{\citenamefont {Dzuba}\ \emph {et~al.}(2011)\citenamefont {Dzuba},
  \citenamefont {Flambaum},\ and\ \citenamefont
  {Harabati}}]{Dzuba-Flambaum-Harabati-ratios:2011}%
  \BibitemOpen
  \bibfield  {author} {\bibinfo {author} {\bibfnamefont {V.~A.}\ \bibnamefont
  {Dzuba}}, \bibinfo {author} {\bibfnamefont {V.~V.}\ \bibnamefont {Flambaum}},
  \ and\ \bibinfo {author} {\bibfnamefont {C.}~\bibnamefont {Harabati}},\
  }\href@noop {} {\bibfield  {journal} {\bibinfo  {journal} {Phys.~Rev.~A}\
  }\textbf {\bibinfo {volume} {84}},\ \bibinfo {pages} {052108} (\bibinfo
  {year} {2011})}\BibitemShut {NoStop}%
\bibitem [{\citenamefont {Gould}\ and\ \citenamefont
  {Jr.}(2014)}]{GouldMunger:2014}%
  \BibitemOpen
  \bibfield  {author} {\bibinfo {author} {\bibfnamefont {H.}~\bibnamefont
  {Gould}}\ and\ \bibinfo {author} {\bibfnamefont {C.~T.~M.}\ \bibnamefont
  {Jr.}},\ }\href@noop {} {\enquote {\bibinfo {title} {Position paper: Electric
  dipole moment experiments},}\ }\bibinfo {howpublished}
  {http://fsnutown.phy.ornl.gov/fsnuweb/index.html} (\bibinfo {year} {28-29 Sep
  2014})\BibitemShut {NoStop}%
\bibitem [{\citenamefont {Swallows}\ \emph
  {et~al.}(2013{\natexlab{a}})\citenamefont {Swallows}, \citenamefont {Loftus},
  \citenamefont {Griffith}, \citenamefont {Heckel},\ and\ \citenamefont
  {Fortson}}]{Swallows:2013}%
  \BibitemOpen
  \bibfield  {author} {\bibinfo {author} {\bibfnamefont {M.~D.}\ \bibnamefont
  {Swallows}}, \bibinfo {author} {\bibfnamefont {T.~H.}\ \bibnamefont
  {Loftus}}, \bibinfo {author} {\bibfnamefont {W.~C.}\ \bibnamefont
  {Griffith}}, \bibinfo {author} {\bibfnamefont {B.~R.}\ \bibnamefont
  {Heckel}}, \ and\ \bibinfo {author} {\bibfnamefont {E.~N.}\ \bibnamefont
  {Fortson}},\ }\href@noop {} {\bibfield  {journal} {\bibinfo  {journal}
  {Phys.~Rev.~A}\ }\textbf {\bibinfo {volume} {87}},\ \bibinfo {pages} {012102}
  (\bibinfo {year} {2013}{\natexlab{a}})}\BibitemShut {NoStop}%
\bibitem [{\citenamefont {Eichler}\ \emph {et~al.}(2002)\citenamefont
  {Eichler}, \citenamefont {Aksenov}, \citenamefont {Belozerov}, \citenamefont
  {Bozhikov}, \citenamefont {Chepigin}, \citenamefont {Dmitriev}, \citenamefont
  {Dressler}, \citenamefont {Ga{\aa}ggeler}, \citenamefont {Gorshkov},
  \citenamefont {Haenssler}, \citenamefont {Itkis}, \citenamefont {Laube},
  \citenamefont {Lebedev}, \citenamefont {Malyshev}, \citenamefont
  {Oganessian}, \citenamefont {Petrushkin}, \citenamefont {Piguet},
  \citenamefont {Rasmussen}, \citenamefont {Shishkin}, \citenamefont {Shutov},
  \citenamefont {Svirikhin}, \citenamefont {Tereshatov}, \citenamefont
  {Vostokin}, \citenamefont {Wegrzecki},\ and\ \citenamefont
  {Yeremin}}]{Cn-112:Nature:2007}%
  \BibitemOpen
  \bibfield  {author} {\bibinfo {author} {\bibfnamefont {R.}~\bibnamefont
  {Eichler}}, \bibinfo {author} {\bibfnamefont {N.~V.}\ \bibnamefont
  {Aksenov}}, \bibinfo {author} {\bibfnamefont {A.~V.}\ \bibnamefont
  {Belozerov}}, \bibinfo {author} {\bibfnamefont {G.~A.}\ \bibnamefont
  {Bozhikov}}, \bibinfo {author} {\bibfnamefont {V.~I.}\ \bibnamefont
  {Chepigin}}, \bibinfo {author} {\bibfnamefont {S.~N.}\ \bibnamefont
  {Dmitriev}}, \bibinfo {author} {\bibfnamefont {R.}~\bibnamefont {Dressler}},
  \bibinfo {author} {\bibfnamefont {H.~W.}\ \bibnamefont {Ga{\aa}ggeler}},
  \bibinfo {author} {\bibfnamefont {V.~A.}\ \bibnamefont {Gorshkov}}, \bibinfo
  {author} {\bibfnamefont {F.}~\bibnamefont {Haenssler}}, \bibinfo {author}
  {\bibfnamefont {M.~G.}\ \bibnamefont {Itkis}}, \bibinfo {author}
  {\bibfnamefont {A.}~\bibnamefont {Laube}}, \bibinfo {author} {\bibfnamefont
  {V.~Y.}\ \bibnamefont {Lebedev}}, \bibinfo {author} {\bibfnamefont {O.~N.}\
  \bibnamefont {Malyshev}}, \bibinfo {author} {\bibfnamefont {Y.~T.}\
  \bibnamefont {Oganessian}}, \bibinfo {author} {\bibfnamefont {O.~V.}\
  \bibnamefont {Petrushkin}}, \bibinfo {author} {\bibfnamefont
  {D.}~\bibnamefont {Piguet}}, \bibinfo {author} {\bibfnamefont
  {P.}~\bibnamefont {Rasmussen}}, \bibinfo {author} {\bibfnamefont {S.~V.}\
  \bibnamefont {Shishkin}}, \bibinfo {author} {\bibfnamefont {A.~V.}\
  \bibnamefont {Shutov}}, \bibinfo {author} {\bibfnamefont {A.~I.}\
  \bibnamefont {Svirikhin}}, \bibinfo {author} {\bibfnamefont {E.~E.}\
  \bibnamefont {Tereshatov}}, \bibinfo {author} {\bibfnamefont {G.~K.}\
  \bibnamefont {Vostokin}}, \bibinfo {author} {\bibfnamefont {M.}~\bibnamefont
  {Wegrzecki}}, \ and\ \bibinfo {author} {\bibfnamefont {A.~V.}\ \bibnamefont
  {Yeremin}},\ }\href@noop {} {\bibfield  {journal} {\bibinfo  {journal}
  {Nature}\ }\textbf {\bibinfo {volume} {447}},\ \bibinfo {pages} {72}
  (\bibinfo {year} {2002})}\BibitemShut {NoStop}%
\bibitem [{\citenamefont {Barber}\ \emph {et~al.}(2009)\citenamefont {Barber},
  \citenamefont {G{\aa}ggeler}, \citenamefont {Karol}, \citenamefont
  {Nakahara}, \citenamefont {Vardaci},\ and\ \citenamefont
  {Vogt}}]{Cn-112:IUPAC:2009}%
  \BibitemOpen
  \bibfield  {author} {\bibinfo {author} {\bibfnamefont {R.~C.}\ \bibnamefont
  {Barber}}, \bibinfo {author} {\bibfnamefont {H.~W.}\ \bibnamefont
  {G{\aa}ggeler}}, \bibinfo {author} {\bibfnamefont {P.~J.}\ \bibnamefont
  {Karol}}, \bibinfo {author} {\bibfnamefont {H.}~\bibnamefont {Nakahara}},
  \bibinfo {author} {\bibfnamefont {E.}~\bibnamefont {Vardaci}}, \ and\
  \bibinfo {author} {\bibfnamefont {E.}~\bibnamefont {Vogt}},\ }\href@noop {}
  {\bibfield  {journal} {\bibinfo  {journal} {Pure~Appl.~Chem.}\ }\textbf
  {\bibinfo {volume} {81}},\ \bibinfo {pages} {1331–1343} (\bibinfo {year}
  {2009})}\BibitemShut {NoStop}%
\bibitem [{\citenamefont {Dyall}\ \emph {et~al.}(1989)\citenamefont {Dyall},
  \citenamefont {Grant}, \citenamefont {Johnson}, \citenamefont {Parpia},\ and\
  \citenamefont {Plummer}}]{grasp89}%
  \BibitemOpen
  \bibfield  {author} {\bibinfo {author} {\bibfnamefont {K.~G.}\ \bibnamefont
  {Dyall}}, \bibinfo {author} {\bibfnamefont {I.~P.}\ \bibnamefont {Grant}},
  \bibinfo {author} {\bibfnamefont {C.~T.}\ \bibnamefont {Johnson}}, \bibinfo
  {author} {\bibfnamefont {F.~A.}\ \bibnamefont {Parpia}}, \ and\ \bibinfo
  {author} {\bibfnamefont {E.~P.}\ \bibnamefont {Plummer}},\ }\href@noop {}
  {\bibfield  {journal} {\bibinfo  {journal} {Comput.~Phys.~Commun.}\ }\textbf
  {\bibinfo {volume} {55}},\ \bibinfo {pages} {425} (\bibinfo {year}
  {1989})}\BibitemShut {NoStop}%
\bibitem [{\citenamefont {J{\"o}nsson}\ \emph {et~al.}(2013)\citenamefont
  {J{\"o}nsson}, \citenamefont {Gaigalas}, \citenamefont {Biero\'n},
  \citenamefont {{Froese Fischer}},\ and\ \citenamefont
  {Grant}}]{grasp2K:2013}%
  \BibitemOpen
  \bibfield  {author} {\bibinfo {author} {\bibfnamefont {P.}~\bibnamefont
  {J{\"o}nsson}}, \bibinfo {author} {\bibfnamefont {G.}~\bibnamefont
  {Gaigalas}}, \bibinfo {author} {\bibfnamefont {J.}~\bibnamefont {Biero\'n}},
  \bibinfo {author} {\bibfnamefont {C.}~\bibnamefont {{Froese Fischer}}}, \
  and\ \bibinfo {author} {\bibfnamefont {I.~P.}\ \bibnamefont {Grant}},\
  }\href@noop {} {\bibfield  {journal} {\bibinfo  {journal}
  {Comput.~Phys.~Commun.}\ }\textbf {\bibinfo {volume} {184}},\ \bibinfo
  {pages} {2197} (\bibinfo {year} {2013})}\BibitemShut {NoStop}%
\bibitem [{\citenamefont {McKenzie}\ \emph {et~al.}(1980)\citenamefont
  {McKenzie}, \citenamefont {Grant},\ and\ \citenamefont
  {Norrington}}]{graspMcKenzie1980}%
  \BibitemOpen
  \bibfield  {author} {\bibinfo {author} {\bibfnamefont {B.~J.}\ \bibnamefont
  {McKenzie}}, \bibinfo {author} {\bibfnamefont {I.~P.}\ \bibnamefont {Grant}},
  \ and\ \bibinfo {author} {\bibfnamefont {P.~H.}\ \bibnamefont {Norrington}},\
  }\href@noop {} {\bibfield  {journal} {\bibinfo  {journal}
  {Comput.~Phys.~Commun.}\ }\textbf {\bibinfo {volume} {21}},\ \bibinfo {pages}
  {233} (\bibinfo {year} {1980})}\BibitemShut {NoStop}%
\bibitem [{\citenamefont {Parpia}\ \emph {et~al.}(1996)\citenamefont {Parpia},
  \citenamefont {{Froese Fischer}},\ and\ \citenamefont {Grant}}]{grasp92}%
  \BibitemOpen
  \bibfield  {author} {\bibinfo {author} {\bibfnamefont {F.~A.}\ \bibnamefont
  {Parpia}}, \bibinfo {author} {\bibfnamefont {C.}~\bibnamefont {{Froese
  Fischer}}}, \ and\ \bibinfo {author} {\bibfnamefont {I.~P.}\ \bibnamefont
  {Grant}},\ }\href@noop {} {\bibfield  {journal} {\bibinfo  {journal}
  {Comput.~Phys.~Commun.}\ }\textbf {\bibinfo {volume} {94}},\ \bibinfo {pages}
  {249} (\bibinfo {year} {1996})}\BibitemShut {NoStop}%
\bibitem [{\citenamefont {Grant}(1994)}]{Grant1994}%
  \BibitemOpen
  \bibfield  {author} {\bibinfo {author} {\bibfnamefont {I.~P.}\ \bibnamefont
  {Grant}},\ }\href@noop {} {\bibfield  {journal} {\bibinfo  {journal}
  {Comput.~Phys.~Commun.}\ }\textbf {\bibinfo {volume} {84}},\ \bibinfo {pages}
  {59} (\bibinfo {year} {1994})}\BibitemShut {NoStop}%
\bibitem [{\citenamefont {J{\"o}nsson}\ \emph {et~al.}(2007)\citenamefont
  {J{\"o}nsson}, \citenamefont {He},\ and\ \citenamefont {{Froese
  Fischer}}}]{grasp2K}%
  \BibitemOpen
  \bibfield  {author} {\bibinfo {author} {\bibfnamefont {P.}~\bibnamefont
  {J{\"o}nsson}}, \bibinfo {author} {\bibfnamefont {X.}~\bibnamefont {He}}, \
  and\ \bibinfo {author} {\bibfnamefont {C.}~\bibnamefont {{Froese Fischer}}},\
  }\href@noop {} {\bibfield  {journal} {\bibinfo  {journal}
  {Comput.~Phys.~Commun.}\ }\textbf {\bibinfo {volume} {176}},\ \bibinfo
  {pages} {597} (\bibinfo {year} {2007})}\BibitemShut {NoStop}%
\bibitem [{\citenamefont {Grant}(2007)}]{GrantBook2007}%
  \BibitemOpen
  \bibfield  {author} {\bibinfo {author} {\bibfnamefont {I.~P.}\ \bibnamefont
  {Grant}},\ }\href@noop {} {\emph {\bibinfo {title} {Relativistic Quantum
  Theory of Atoms and Molecules: Theory and Computation}}}\ (\bibinfo
  {publisher} {Springer},\ \bibinfo {address} {New York},\ \bibinfo {year}
  {2007})\BibitemShut {NoStop}%
\bibitem [{\citenamefont {Rad{\v z}i{\= u}t{\.e}}\ \emph
  {et~al.}(2014)\citenamefont {Rad{\v z}i{\= u}t{\.e}}, \citenamefont
  {Gaigalas}, \citenamefont {J{\"o}nsson},\ and\ \citenamefont
  {Biero\'n}}]{RadziuteRaHgYbEDM2014}%
  \BibitemOpen
  \bibfield  {author} {\bibinfo {author} {\bibfnamefont {L.}~\bibnamefont
  {Rad{\v z}i{\= u}t{\.e}}}, \bibinfo {author} {\bibfnamefont {G.}~\bibnamefont
  {Gaigalas}}, \bibinfo {author} {\bibfnamefont {P.}~\bibnamefont
  {J{\"o}nsson}}, \ and\ \bibinfo {author} {\bibfnamefont {J.}~\bibnamefont
  {Biero\'n}},\ }\href@noop {} {\bibfield  {journal} {\bibinfo  {journal}
  {Phys.~Rev.~A}\ }\textbf {\bibinfo {volume} {90}},\ \bibinfo {pages} {012528}
  (\bibinfo {year} {2014})}\BibitemShut {NoStop}%
\bibitem [{\citenamefont {Biero{\'n}}\ \emph {et~al.}(2009)\citenamefont
  {Biero{\'n}}, \citenamefont {{Froese Fischer}}, \citenamefont {Indelicato},
  \citenamefont {J{\"o}nsson},\ and\ \citenamefont
  {Pyykk{\"o}}}]{BieronAu2009}%
  \BibitemOpen
  \bibfield  {author} {\bibinfo {author} {\bibfnamefont {J.}~\bibnamefont
  {Biero{\'n}}}, \bibinfo {author} {\bibfnamefont {C.}~\bibnamefont {{Froese
  Fischer}}}, \bibinfo {author} {\bibfnamefont {P.}~\bibnamefont {Indelicato}},
  \bibinfo {author} {\bibfnamefont {P.}~\bibnamefont {J{\"o}nsson}}, \ and\
  \bibinfo {author} {\bibfnamefont {P.}~\bibnamefont {Pyykk{\"o}}},\
  }\href@noop {} {\bibfield  {journal} {\bibinfo  {journal} {Phys.~Rev.~A}\
  }\textbf {\bibinfo {volume} {79}},\ \bibinfo {pages} {052502} (\bibinfo
  {year} {2009})}\BibitemShut {NoStop}%
\bibitem [{\citenamefont {Biero{\'n}}\ \emph {et~al.}(2015)\citenamefont
  {Biero{\'n}}, \citenamefont {{Froese Fischer}}, \citenamefont {Fritzsche},
  \citenamefont {Gaigalas}, \citenamefont {Grant}, \citenamefont {Indelicato},
  \citenamefont {J{\"o}nsson},\ and\ \citenamefont
  {Pyykk{\"o}}}]{Bieron:e-N:2015}%
  \BibitemOpen
  \bibfield  {author} {\bibinfo {author} {\bibfnamefont {J.}~\bibnamefont
  {Biero{\'n}}}, \bibinfo {author} {\bibfnamefont {C.}~\bibnamefont {{Froese
  Fischer}}}, \bibinfo {author} {\bibfnamefont {S.}~\bibnamefont {Fritzsche}},
  \bibinfo {author} {\bibfnamefont {G.}~\bibnamefont {Gaigalas}}, \bibinfo
  {author} {\bibfnamefont {I.~P.}\ \bibnamefont {Grant}}, \bibinfo {author}
  {\bibfnamefont {P.}~\bibnamefont {Indelicato}}, \bibinfo {author}
  {\bibfnamefont {P.}~\bibnamefont {J{\"o}nsson}}, \ and\ \bibinfo {author}
  {\bibfnamefont {P.}~\bibnamefont {Pyykk{\"o}}},\ }\href@noop {} {\bibfield
  {journal} {\bibinfo  {journal} {Phys.~Scr.}\ }\textbf {\bibinfo {volume}
  {90}},\ \bibinfo {pages} {054011} (\bibinfo {year} {2015})}\BibitemShut
  {NoStop}%
\bibitem [{\citenamefont {Rad{\v z}i{\= u}t{\.e}}\ \emph
  {et~al.}(2015)\citenamefont {Rad{\v z}i{\= u}t{\.e}}, \citenamefont
  {Gaigalas}, \citenamefont {Kato}, \citenamefont {J{\"o}nsson}, \citenamefont
  {Rynkun}, \citenamefont {Ku{\v c}as}, \citenamefont {Jonauskas},\ and\
  \citenamefont {Matulianec}}]{Radziute2015}%
  \BibitemOpen
  \bibfield  {author} {\bibinfo {author} {\bibfnamefont {L.}~\bibnamefont
  {Rad{\v z}i{\= u}t{\.e}}}, \bibinfo {author} {\bibfnamefont {G.}~\bibnamefont
  {Gaigalas}}, \bibinfo {author} {\bibfnamefont {D.}~\bibnamefont {Kato}},
  \bibinfo {author} {\bibfnamefont {P.}~\bibnamefont {J{\"o}nsson}}, \bibinfo
  {author} {\bibfnamefont {P.}~\bibnamefont {Rynkun}}, \bibinfo {author}
  {\bibfnamefont {S.}~\bibnamefont {Ku{\v c}as}}, \bibinfo {author}
  {\bibfnamefont {V.}~\bibnamefont {Jonauskas}}, \ and\ \bibinfo {author}
  {\bibfnamefont {R.}~\bibnamefont {Matulianec}},\ }\href@noop {} {\bibfield
  {journal} {\bibinfo  {journal} {J Quant Spectrosc Radiat Transf}\ }\textbf
  {\bibinfo {volume} {152}},\ \bibinfo {pages} {94} (\bibinfo {year}
  {2015})}\BibitemShut {NoStop}%
\bibitem [{\citenamefont {Gaigalas}\ \emph {et~al.}(1997)\citenamefont
  {Gaigalas}, \citenamefont {Rudzikas},\ and\ \citenamefont {{Froese
  Fischer}}}]{Gaigalas1997}%
  \BibitemOpen
  \bibfield  {author} {\bibinfo {author} {\bibfnamefont {G.}~\bibnamefont
  {Gaigalas}}, \bibinfo {author} {\bibfnamefont {Z.}~\bibnamefont {Rudzikas}},
  \ and\ \bibinfo {author} {\bibfnamefont {C.}~\bibnamefont {{Froese
  Fischer}}},\ }\href@noop {} {\bibfield  {journal} {\bibinfo  {journal}
  {J.~Phys.~B:~At.~Mol.~Opt.~Phys.}\ }\textbf {\bibinfo {volume} {30}},\
  \bibinfo {pages} {3747} (\bibinfo {year} {1997})}\BibitemShut {NoStop}%
\bibitem [{\citenamefont {Cowan}\ and\ \citenamefont
  {Hansen}(1981)}]{CowanHansen:1981}%
  \BibitemOpen
  \bibfield  {author} {\bibinfo {author} {\bibfnamefont {R.~D.}\ \bibnamefont
  {Cowan}}\ and\ \bibinfo {author} {\bibfnamefont {J.~E.}\ \bibnamefont
  {Hansen}},\ }\href@noop {} {\bibfield  {journal} {\bibinfo  {journal}
  {J.~Opt.~Soc.~Am.}\ }\textbf {\bibinfo {volume} {71}},\ \bibinfo {pages} {60}
  (\bibinfo {year} {1981})}\BibitemShut {NoStop}%
\bibitem [{\citenamefont {Syty}\ \emph {et~al.}(2016)\citenamefont {Syty},
  \citenamefont {Sienkiewicz}, \citenamefont {Rad{\v z}i{\= u}t{\.e}},
  \citenamefont {Gaigalas},\ and\ \citenamefont
  {Biero\'n}}]{edm-continuum-work-in-progress}%
  \BibitemOpen
  \bibfield  {author} {\bibinfo {author} {\bibfnamefont {P.}~\bibnamefont
  {Syty}}, \bibinfo {author} {\bibfnamefont {J.}~\bibnamefont {Sienkiewicz}},
  \bibinfo {author} {\bibfnamefont {L.}~\bibnamefont {Rad{\v z}i{\= u}t{\.e}}},
  \bibinfo {author} {\bibfnamefont {G.}~\bibnamefont {Gaigalas}}, \ and\
  \bibinfo {author} {\bibfnamefont {J.}~\bibnamefont {Biero\'n}},\ }\href@noop
  {} {} (\bibinfo {year} {2016}),\ \bibinfo {note} {work in
  progress}\BibitemShut {NoStop}%
\bibitem [{\citenamefont {Fano}\ and\ \citenamefont
  {Cooper}(1965)}]{FanoCooper:1965}%
  \BibitemOpen
  \bibfield  {author} {\bibinfo {author} {\bibfnamefont {U.}~\bibnamefont
  {Fano}}\ and\ \bibinfo {author} {\bibfnamefont {J.~W.}\ \bibnamefont
  {Cooper}},\ }\href@noop {} {\bibfield  {journal} {\bibinfo  {journal}
  {Rev.~Mod.~Phys.}\ }\textbf {\bibinfo {volume} {40}},\ \bibinfo {pages}
  {441–507} (\bibinfo {year} {1965})}\BibitemShut {NoStop}%
\bibitem [{\citenamefont {Dzuba}\ \emph {et~al.}(2009)\citenamefont {Dzuba},
  \citenamefont {Flambaum},\ and\ \citenamefont {Porsev}}]{DzubaFlambaum:2009}%
  \BibitemOpen
  \bibfield  {author} {\bibinfo {author} {\bibfnamefont {V.~A.}\ \bibnamefont
  {Dzuba}}, \bibinfo {author} {\bibfnamefont {V.~V.}\ \bibnamefont {Flambaum}},
  \ and\ \bibinfo {author} {\bibfnamefont {S.~G.}\ \bibnamefont {Porsev}},\
  }\href@noop {} {\bibfield  {journal} {\bibinfo  {journal} {Phys.~Rev.~A}\
  }\textbf {\bibinfo {volume} {80}},\ \bibinfo {pages} {032120} (\bibinfo
  {year} {2009})}\BibitemShut {NoStop}%
\bibitem [{\citenamefont {M{\aa}rtensson-Pendrill}(1985)}]{Martensson:1985}%
  \BibitemOpen
  \bibfield  {author} {\bibinfo {author} {\bibfnamefont {A.-M.}\ \bibnamefont
  {M{\aa}rtensson-Pendrill}},\ }\href@noop {} {\bibfield  {journal} {\bibinfo
  {journal} {Phys.~Rev.~Lett.}\ }\textbf {\bibinfo {volume} {54}},\ \bibinfo
  {pages} {1153} (\bibinfo {year} {1985})}\BibitemShut {NoStop}%
\bibitem [{\citenamefont {Latha}\ \emph {et~al.}(2008)\citenamefont {Latha},
  \citenamefont {Angom}, \citenamefont {Chaudhuri}, \citenamefont {Das},\ and\
  \citenamefont {Mukherjee}}]{Latha:2008}%
  \BibitemOpen
  \bibfield  {author} {\bibinfo {author} {\bibfnamefont {K.~V.~P.}\
  \bibnamefont {Latha}}, \bibinfo {author} {\bibfnamefont {D.}~\bibnamefont
  {Angom}}, \bibinfo {author} {\bibfnamefont {R.~J.}\ \bibnamefont
  {Chaudhuri}}, \bibinfo {author} {\bibfnamefont {B.~P.}\ \bibnamefont {Das}},
  \ and\ \bibinfo {author} {\bibfnamefont {D.}~\bibnamefont {Mukherjee}},\
  }\href@noop {} {\bibfield  {journal} {\bibinfo  {journal}
  {J.~Phys.~B:~At.~Mol.~Opt.~Phys.}\ }\textbf {\bibinfo {volume} {41}},\
  \bibinfo {pages} {035005} (\bibinfo {year} {2008})}\BibitemShut {NoStop}%
\bibitem [{\citenamefont {Latha}\ \emph {et~al.}(2009)\citenamefont {Latha},
  \citenamefont {Angom}, \citenamefont {Das},\ and\ \citenamefont
  {Mukherjee}}]{Latha:2009}%
  \BibitemOpen
  \bibfield  {author} {\bibinfo {author} {\bibfnamefont {K.~V.~P.}\
  \bibnamefont {Latha}}, \bibinfo {author} {\bibfnamefont {D.}~\bibnamefont
  {Angom}}, \bibinfo {author} {\bibfnamefont {B.~P.}\ \bibnamefont {Das}}, \
  and\ \bibinfo {author} {\bibfnamefont {D.}~\bibnamefont {Mukherjee}},\
  }\href@noop {} {\bibfield  {journal} {\bibinfo  {journal} {Phys.~Rev.~Lett.}\
  }\textbf {\bibinfo {volume} {103}},\ \bibinfo {pages} {083001} (\bibinfo
  {year} {2009})}\BibitemShut {NoStop}%
\bibitem [{\citenamefont {Dzuba}\ \emph {et~al.}(2002)\citenamefont {Dzuba},
  \citenamefont {Flambaum}, \citenamefont {Ginges},\ and\ \citenamefont
  {Kozlov}}]{Dzuba:2002}%
  \BibitemOpen
  \bibfield  {author} {\bibinfo {author} {\bibfnamefont {V.~A.}\ \bibnamefont
  {Dzuba}}, \bibinfo {author} {\bibfnamefont {V.~V.}\ \bibnamefont {Flambaum}},
  \bibinfo {author} {\bibfnamefont {J.~S.~M.}\ \bibnamefont {Ginges}}, \ and\
  \bibinfo {author} {\bibfnamefont {M.~G.}\ \bibnamefont {Kozlov}},\
  }\href@noop {} {\bibfield  {journal} {\bibinfo  {journal} {Phys.~Rev.~A}\
  }\textbf {\bibinfo {volume} {66}},\ \bibinfo {pages} {012111} (\bibinfo
  {year} {2002})}\BibitemShut {NoStop}%
\bibitem [{\citenamefont {Dzuba}\ \emph {et~al.}(2007)\citenamefont {Dzuba},
  \citenamefont {Flambaum},\ and\ \citenamefont
  {Ginges}}]{DzubaFlambaum:2007:76}%
  \BibitemOpen
  \bibfield  {author} {\bibinfo {author} {\bibfnamefont {V.~A.}\ \bibnamefont
  {Dzuba}}, \bibinfo {author} {\bibfnamefont {V.~V.}\ \bibnamefont {Flambaum}},
  \ and\ \bibinfo {author} {\bibfnamefont {J.~S.~M.}\ \bibnamefont {Ginges}},\
  }\href@noop {} {\bibfield  {journal} {\bibinfo  {journal} {Phys.~Rev.~A}\
  }\textbf {\bibinfo {volume} {76}},\ \bibinfo {pages} {034501} (\bibinfo
  {year} {2007})}\BibitemShut {NoStop}%
\bibitem [{\citenamefont {M{\aa}rtensson-Pendrill}\ and\ \citenamefont
  {{\"O}ster}(1987)}]{Martensson:1987}%
  \BibitemOpen
  \bibfield  {author} {\bibinfo {author} {\bibfnamefont {A.-M.}\ \bibnamefont
  {M{\aa}rtensson-Pendrill}}\ and\ \bibinfo {author} {\bibfnamefont
  {P.}~\bibnamefont {{\"O}ster}},\ }\href@noop {} {\bibfield  {journal}
  {\bibinfo  {journal} {Phys.~Scr.}\ }\textbf {\bibinfo {volume} {36}},\
  \bibinfo {pages} {444} (\bibinfo {year} {1987})}\BibitemShut {NoStop}%
\bibitem [{\citenamefont {Kramida}\ \emph {et~al.}()\citenamefont {Kramida},
  \citenamefont {{Yu.~Ralchenko}}, \citenamefont {Reader},\ and\ \citenamefont
  {{NIST ASD Team}}}]{NIST_ASD}%
  \BibitemOpen
  \bibfield  {author} {\bibinfo {author} {\bibfnamefont {A.}~\bibnamefont
  {Kramida}}, \bibinfo {author} {\bibnamefont {{Yu.~Ralchenko}}}, \bibinfo
  {author} {\bibfnamefont {J.}~\bibnamefont {Reader}}, \ and\ \bibinfo {author}
  {\bibnamefont {{NIST ASD Team}}},\ }\href@noop {} {}\bibinfo {howpublished}
  {\url{http://physics.nist.gov/asd}}\BibitemShut {NoStop}%
\bibitem [{\citenamefont {Gaigalas}\ \emph {et~al.}(2003)\citenamefont
  {Gaigalas}, \citenamefont {{\v Z}alandauskas},\ and\ \citenamefont
  {Rudzikas}}]{Transformation}%
  \BibitemOpen
  \bibfield  {author} {\bibinfo {author} {\bibfnamefont {G.}~\bibnamefont
  {Gaigalas}}, \bibinfo {author} {\bibfnamefont {T.}~\bibnamefont {{\v
  Z}alandauskas}}, \ and\ \bibinfo {author} {\bibfnamefont {Z.}~\bibnamefont
  {Rudzikas}},\ }\href@noop {} {\bibfield  {journal} {\bibinfo  {journal}
  {Atomic~Data~Nucl.~Data~Tab.}\ }\textbf {\bibinfo {volume} {84}},\ \bibinfo
  {pages} {99} (\bibinfo {year} {2003})}\BibitemShut {NoStop}%
\bibitem [{\citenamefont {Gaigalas}\ \emph {et~al.}(2004)\citenamefont
  {Gaigalas}, \citenamefont {{\v Z}alandauskas},\ and\ \citenamefont
  {Fritzsche}}]{LSJ2}%
  \BibitemOpen
  \bibfield  {author} {\bibinfo {author} {\bibfnamefont {G.}~\bibnamefont
  {Gaigalas}}, \bibinfo {author} {\bibfnamefont {T.}~\bibnamefont {{\v
  Z}alandauskas}}, \ and\ \bibinfo {author} {\bibfnamefont {S.}~\bibnamefont
  {Fritzsche}},\ }\href@noop {} {\bibfield  {journal} {\bibinfo  {journal}
  {Comput.~Phys.~Commun.}\ }\textbf {\bibinfo {volume} {157}},\ \bibinfo
  {pages} {239} (\bibinfo {year} {2004})}\BibitemShut {NoStop}%
\bibitem [{\citenamefont {Yu}\ \emph {et~al.}(2007)\citenamefont {Yu},
  \citenamefont {Li}, \citenamefont {Dong}, \citenamefont {Ding}, \citenamefont
  {Fritzsche},\ and\ \citenamefont {Fricke}}]{Fritzsche:2007:2}%
  \BibitemOpen
  \bibfield  {author} {\bibinfo {author} {\bibfnamefont {Y.~J.}\ \bibnamefont
  {Yu}}, \bibinfo {author} {\bibfnamefont {J.~G.}\ \bibnamefont {Li}}, \bibinfo
  {author} {\bibfnamefont {C.~Z.}\ \bibnamefont {Dong}}, \bibinfo {author}
  {\bibfnamefont {X.~B.}\ \bibnamefont {Ding}}, \bibinfo {author}
  {\bibfnamefont {S.}~\bibnamefont {Fritzsche}}, \ and\ \bibinfo {author}
  {\bibfnamefont {B.}~\bibnamefont {Fricke}},\ }\href@noop {} {\bibfield
  {journal} {\bibinfo  {journal} {Eur.~Phys.~J.~D}\ }\textbf {\bibinfo {volume}
  {44}},\ \bibinfo {pages} {51} (\bibinfo {year} {2007})}\BibitemShut {NoStop}%
\bibitem [{\citenamefont {Dinh}\ \emph {et~al.}(2008)\citenamefont {Dinh},
  \citenamefont {Dzuba},\ and\ \citenamefont {Flambaum}}]{Dinh:Dzuba:2008}%
  \BibitemOpen
  \bibfield  {author} {\bibinfo {author} {\bibfnamefont {T.~H.}\ \bibnamefont
  {Dinh}}, \bibinfo {author} {\bibfnamefont {V.~A.}\ \bibnamefont {Dzuba}}, \
  and\ \bibinfo {author} {\bibfnamefont {V.~V.}\ \bibnamefont {Flambaum}},\
  }\href@noop {} {\bibfield  {journal} {\bibinfo  {journal} {Phys.~Rev.~A}\
  }\textbf {\bibinfo {volume} {78}},\ \bibinfo {pages} {062502} (\bibinfo
  {year} {2008})}\BibitemShut {NoStop}%
\bibitem [{\citenamefont {Smola{\'n}czuk}(1997)}]{Smolanczuk:1997}%
  \BibitemOpen
  \bibfield  {author} {\bibinfo {author} {\bibfnamefont {R.}~\bibnamefont
  {Smola{\'n}czuk}},\ }\href@noop {} {\bibfield  {journal} {\bibinfo  {journal}
  {Phys.~Rev.~C}\ }\textbf {\bibinfo {volume} {56}},\ \bibinfo {pages} {812}
  (\bibinfo {year} {1997})}\BibitemShut {NoStop}%
\bibitem [{\citenamefont {Zagrebaev}\ \emph {et~al.}(2013)\citenamefont
  {Zagrebaev}, \citenamefont {Karpov},\ and\ \citenamefont
  {Greiner}}]{Zagrebaev:2013}%
  \BibitemOpen
  \bibfield  {author} {\bibinfo {author} {\bibfnamefont {V.}~\bibnamefont
  {Zagrebaev}}, \bibinfo {author} {\bibfnamefont {A.}~\bibnamefont {Karpov}}, \
  and\ \bibinfo {author} {\bibfnamefont {W.}~\bibnamefont {Greiner}},\
  }\href@noop {} {\bibfield  {journal} {\bibinfo  {journal}
  {J.~Phys.:~Conf.~Ser.}\ }\textbf {\bibinfo {volume} {420}},\ \bibinfo {pages}
  {012001} (\bibinfo {year} {2013})}\BibitemShut {NoStop}%
\bibitem [{nud()}]{nudat2}%
  \BibitemOpen
  \href@noop {} {}\bibinfo {note}
  {\url{http://www.nndc.bnl.gov/nudat2}}\BibitemShut {NoStop}%
\bibitem [{\citenamefont {Graner}\ \emph {et~al.}(2016)\citenamefont {Graner},
  \citenamefont {Chen}, \citenamefont {Lindahl},\ and\ \citenamefont
  {Heckel}}]{Graner:2016}%
  \BibitemOpen
  \bibfield  {author} {\bibinfo {author} {\bibfnamefont {B.}~\bibnamefont
  {Graner}}, \bibinfo {author} {\bibfnamefont {Y.}~\bibnamefont {Chen}},
  \bibinfo {author} {\bibfnamefont {E.~G.}\ \bibnamefont {Lindahl}}, \ and\
  \bibinfo {author} {\bibfnamefont {B.~R.}\ \bibnamefont {Heckel}},\
  }\href@noop {} {\  (\bibinfo {year} {2016})},\ \Eprint
  {http://arxiv.org/abs/1601.04339} {arXiv:1601.04339 [physics.atom-ph]}
  \BibitemShut {NoStop}%
\bibitem [{\citenamefont {Fricke}\ \emph {et~al.}(1971)\citenamefont {Fricke},
  \citenamefont {Greiner},\ and\ \citenamefont {Waber}}]{Fricke:Z-172:1970}%
  \BibitemOpen
  \bibfield  {author} {\bibinfo {author} {\bibfnamefont {B.}~\bibnamefont
  {Fricke}}, \bibinfo {author} {\bibfnamefont {W.}~\bibnamefont {Greiner}}, \
  and\ \bibinfo {author} {\bibfnamefont {J.~T.}\ \bibnamefont {Waber}},\
  }\href@noop {} {\bibfield  {journal} {\bibinfo  {journal}
  {Theoret.~Chim.~Acta}\ }\textbf {\bibinfo {volume} {21}},\ \bibinfo {pages}
  {235} (\bibinfo {year} {1971})}\BibitemShut {NoStop}%
\bibitem [{\citenamefont {Penneman}\ \emph {et~al.}(1971)\citenamefont
  {Penneman}, \citenamefont {Mann},\ and\ \citenamefont
  {J{\o}rgensen}}]{Penneman:Z-164:1971}%
  \BibitemOpen
  \bibfield  {author} {\bibinfo {author} {\bibfnamefont {R.~A.}\ \bibnamefont
  {Penneman}}, \bibinfo {author} {\bibfnamefont {J.~B.}\ \bibnamefont {Mann}},
  \ and\ \bibinfo {author} {\bibfnamefont {C.~K.}\ \bibnamefont
  {J{\o}rgensen}},\ }\href@noop {} {\bibfield  {journal} {\bibinfo  {journal}
  {Chem.~Phys.~Lett.}\ }\textbf {\bibinfo {volume} {8}},\ \bibinfo {pages}
  {321} (\bibinfo {year} {1971})}\BibitemShut {NoStop}%
\bibitem [{\citenamefont {Pyykk{\"o}}(2011)}]{Pyykko:Z-172:2011}%
  \BibitemOpen
  \bibfield  {author} {\bibinfo {author} {\bibfnamefont {P.}~\bibnamefont
  {Pyykk{\"o}}},\ }\href@noop {} {\bibfield  {journal} {\bibinfo  {journal}
  {Phys.~Chem.~Chem.~Phys.}\ }\textbf {\bibinfo {volume} {13}},\ \bibinfo
  {pages} {161} (\bibinfo {year} {2011})}\BibitemShut {NoStop}%
\bibitem [{\citenamefont {{Froese Fischer}}\ \emph {et~al.}(1997)\citenamefont
  {{Froese Fischer}}, \citenamefont {Brage},\ and\ \citenamefont
  {J{\"o}nsson}}]{FBJbook}%
  \BibitemOpen
  \bibfield  {author} {\bibinfo {author} {\bibfnamefont {C.}~\bibnamefont
  {{Froese Fischer}}}, \bibinfo {author} {\bibfnamefont {T.}~\bibnamefont
  {Brage}}, \ and\ \bibinfo {author} {\bibfnamefont {P.}~\bibnamefont
  {J{\"o}nsson}},\ }\href@noop {} {\emph {\bibinfo {title} {Computational
  Atomic Structure. An MCHF Approach}}}\ (\bibinfo  {publisher} {Institute of
  Physics Publishing},\ \bibinfo {address} {Bristol and Philadelphia},\
  \bibinfo {year} {1997})\BibitemShut {NoStop}%
\bibitem [{\citenamefont {Layzer}(1959)}]{Layzer:1959}%
  \BibitemOpen
  \bibfield  {author} {\bibinfo {author} {\bibfnamefont {D.}~\bibnamefont
  {Layzer}},\ }\href@noop {} {\bibfield  {journal} {\bibinfo  {journal}
  {Ann.~Phys}\ }\textbf {\bibinfo {volume} {8}},\ \bibinfo {pages} {271}
  (\bibinfo {year} {1959})}\BibitemShut {NoStop}%
\bibitem [{\citenamefont {Layzer}\ \emph {et~al.}(1964)\citenamefont {Layzer},
  \citenamefont {Hor\'{a}k}, \citenamefont {Lewis},\ and\ \citenamefont
  {Thompson}}]{Layzer:1964}%
  \BibitemOpen
  \bibfield  {author} {\bibinfo {author} {\bibfnamefont {D.}~\bibnamefont
  {Layzer}}, \bibinfo {author} {\bibfnamefont {Z.}~\bibnamefont {Hor\'{a}k}},
  \bibinfo {author} {\bibfnamefont {M.~N.}\ \bibnamefont {Lewis}}, \ and\
  \bibinfo {author} {\bibfnamefont {D.~P.}\ \bibnamefont {Thompson}},\
  }\href@noop {} {\bibfield  {journal} {\bibinfo  {journal} {Ann.~Phys}\
  }\textbf {\bibinfo {volume} {29}},\ \bibinfo {pages} {101} (\bibinfo {year}
  {1964})}\BibitemShut {NoStop}%
\bibitem [{\citenamefont {Edl{\'e}n}(1964)}]{Edlen:1964}%
  \BibitemOpen
  \bibfield  {author} {\bibinfo {author} {\bibfnamefont {B.}~\bibnamefont
  {Edl{\'e}n}},\ }\enquote {\bibinfo {title} {Handbuch der physik},}\ \
  (\bibinfo  {publisher} {Springer},\ \bibinfo {address} {Berlin},\ \bibinfo
  {year} {1964})\ p.~\bibinfo {pages} {80}\BibitemShut {NoStop}%
\bibitem [{\citenamefont {Cowan}(1981)}]{Cowan.book}%
  \BibitemOpen
  \bibfield  {author} {\bibinfo {author} {\bibfnamefont {R.~D.}\ \bibnamefont
  {Cowan}},\ }\href@noop {} {\emph {\bibinfo {title} {The {T}heory of {A}tomic
  {S}tructure and {S}pectra}}}\ (\bibinfo  {publisher} {University of
  {C}alifornia Press, {L}td},\ \bibinfo {address} {Berkeley},\ \bibinfo {year}
  {1981})\BibitemShut {NoStop}%
\bibitem [{\citenamefont {Wiese}\ and\ \citenamefont
  {Weiss}(1968)}]{WieseWeiss:1968}%
  \BibitemOpen
  \bibfield  {author} {\bibinfo {author} {\bibfnamefont {W.~L.}\ \bibnamefont
  {Wiese}}\ and\ \bibinfo {author} {\bibfnamefont {A.~W.}\ \bibnamefont
  {Weiss}},\ }\href@noop {} {\bibfield  {journal} {\bibinfo  {journal}
  {Phys.~Rev.}\ }\textbf {\bibinfo {volume} {175}},\ \bibinfo {pages} {50}
  (\bibinfo {year} {1968})}\BibitemShut {NoStop}%
\bibitem [{\citenamefont {Bouchiat}\ and\ \citenamefont
  {Bouchiat}(1974{\natexlab{a}})}]{BouchiatBouchiat:1974}%
  \BibitemOpen
  \bibfield  {author} {\bibinfo {author} {\bibfnamefont {M.~A.}\ \bibnamefont
  {Bouchiat}}\ and\ \bibinfo {author} {\bibfnamefont {C.}~\bibnamefont
  {Bouchiat}},\ }\href@noop {} {\bibfield  {journal} {\bibinfo  {journal}
  {J.~de Physique}\ }\textbf {\bibinfo {volume} {35}},\ \bibinfo {pages} {899}
  (\bibinfo {year} {1974}{\natexlab{a}})}\BibitemShut {NoStop}%
\bibitem [{\citenamefont {Bouchiat}\ and\ \citenamefont
  {Bouchiat}(1974{\natexlab{b}})}]{Bouchiat-Bouchiat-Phys.Lett.B48-111-1974.pdf}%
  \BibitemOpen
  \bibfield  {author} {\bibinfo {author} {\bibfnamefont {M.~A.}\ \bibnamefont
  {Bouchiat}}\ and\ \bibinfo {author} {\bibfnamefont {C.~C.}\ \bibnamefont
  {Bouchiat}},\ }\href@noop {} {\bibfield  {journal} {\bibinfo  {journal}
  {Phys.Lett.}\ }\textbf {\bibinfo {volume} {48B}},\ \bibinfo {pages} {111}
  (\bibinfo {year} {1974}{\natexlab{b}})}\BibitemShut {NoStop}%
\bibitem [{\citenamefont {Bouchiat}\ and\ \citenamefont
  {Bouchiat}(1975)}]{BouchiatBouchiat:1975}%
  \BibitemOpen
  \bibfield  {author} {\bibinfo {author} {\bibfnamefont {M.~A.}\ \bibnamefont
  {Bouchiat}}\ and\ \bibinfo {author} {\bibfnamefont {C.}~\bibnamefont
  {Bouchiat}},\ }\href@noop {} {\bibfield  {journal} {\bibinfo  {journal}
  {J.~de Physique}\ }\textbf {\bibinfo {volume} {36}},\ \bibinfo {pages} {493}
  (\bibinfo {year} {1975})}\BibitemShut {NoStop}%
\bibitem [{\citenamefont {Sushkov}\ \emph {et~al.}(1984)\citenamefont
  {Sushkov}, \citenamefont {Flambaum},\ and\ \citenamefont
  {Khriplovich}}]{SushkovFlambaumKhriplovich:1984}%
  \BibitemOpen
  \bibfield  {author} {\bibinfo {author} {\bibfnamefont {O.~P.}\ \bibnamefont
  {Sushkov}}, \bibinfo {author} {\bibfnamefont {V.~V.}\ \bibnamefont
  {Flambaum}}, \ and\ \bibinfo {author} {\bibfnamefont {I.~B.}\ \bibnamefont
  {Khriplovich}},\ }\href@noop {} {\bibfield  {journal} {\bibinfo  {journal}
  {Zh.~Eksp.~Teor.~Fiz.}\ }\textbf {\bibinfo {volume} {87}},\ \bibinfo {pages}
  {1521} (\bibinfo {year} {1984})}\BibitemShut {NoStop}%
\bibitem [{\citenamefont {Khriplovich}(1991)}]{Khriplovich:1991}%
  \BibitemOpen
  \bibfield  {author} {\bibinfo {author} {\bibfnamefont {I.~B.}\ \bibnamefont
  {Khriplovich}},\ }\href@noop {} {\emph {\bibinfo {title} {Parity
  Nonconservation in Atomic Phenomena}}}\ (\bibinfo  {publisher} {Gordon and
  Breach},\ \bibinfo {address} {New York},\ \bibinfo {year} {1991})\BibitemShut
  {NoStop}%
\bibitem [{\citenamefont {Flambaum}\ and\ \citenamefont
  {Ginges}(2002)}]{FlambaumGinges:2002}%
  \BibitemOpen
  \bibfield  {author} {\bibinfo {author} {\bibfnamefont {V.~V.}\ \bibnamefont
  {Flambaum}}\ and\ \bibinfo {author} {\bibfnamefont {J.~S.~M.}\ \bibnamefont
  {Ginges}},\ }\href@noop {} {\bibfield  {journal} {\bibinfo  {journal}
  {Phys.~Rev.~A}\ }\textbf {\bibinfo {volume} {65}},\ \bibinfo {pages} {032113}
  (\bibinfo {year} {2002})}\BibitemShut {NoStop}%
\bibitem [{\citenamefont {Rohlf}(1994)}]{Rohlf:1994}%
  \BibitemOpen
  \bibfield  {author} {\bibinfo {author} {\bibfnamefont {J.~W.}\ \bibnamefont
  {Rohlf}},\ }\href@noop {} {\emph {\bibinfo {title} {Modern Physics from
  $\alpha$ to Z$^0$}}}\ (\bibinfo  {publisher} {Wiley},\ \bibinfo {year}
  {1994})\BibitemShut {NoStop}%
\bibitem [{\citenamefont {Greiner}(1990)}]{GreinerBookRQM}%
  \BibitemOpen
  \bibfield  {author} {\bibinfo {author} {\bibfnamefont {W.}~\bibnamefont
  {Greiner}},\ }\href@noop {} {\emph {\bibinfo {title} {Relativistic Quantum
  Mechanics --- Wave Equations}}}\ (\bibinfo  {publisher} {Springer-Verlag},\
  \bibinfo {address} {Berlin Heidelberg},\ \bibinfo {year} {1990})\BibitemShut
  {NoStop}%
\bibitem [{\citenamefont {Greiner}(1994)}]{GreinerBookQE}%
  \BibitemOpen
  \bibfield  {author} {\bibinfo {author} {\bibfnamefont {W.}~\bibnamefont
  {Greiner}},\ }\href@noop {} {\emph {\bibinfo {title} {Quantum
  Electrodynamics}}}\ (\bibinfo  {publisher} {Springer-Verlag},\ \bibinfo
  {address} {Berlin Heidelberg},\ \bibinfo {year} {1994})\BibitemShut {NoStop}%
\bibitem [{\citenamefont {Indelicato}\ \emph {et~al.}(2011)\citenamefont
  {Indelicato}, \citenamefont {Biero\'n},\ and\ \citenamefont
  {J{\"o}nsson}}]{Indelicato:Z-173:2011}%
  \BibitemOpen
  \bibfield  {author} {\bibinfo {author} {\bibfnamefont {P.}~\bibnamefont
  {Indelicato}}, \bibinfo {author} {\bibfnamefont {J.}~\bibnamefont
  {Biero\'n}}, \ and\ \bibinfo {author} {\bibfnamefont {P.}~\bibnamefont
  {J{\"o}nsson}},\ }\href@noop {} {\bibfield  {journal} {\bibinfo  {journal}
  {Theor.~Chem.~Acc.}\ }\textbf {\bibinfo {volume} {129}},\ \bibinfo {pages}
  {495} (\bibinfo {year} {2011})}\BibitemShut {NoStop}%
\bibitem [{\citenamefont {Goidenko}\ \emph {et~al.}(2007)\citenamefont
  {Goidenko}, \citenamefont {Tupitsyn},\ and\ \citenamefont
  {Plunien}}]{Goidenko:QED-112:2007}%
  \BibitemOpen
  \bibfield  {author} {\bibinfo {author} {\bibfnamefont {I.}~\bibnamefont
  {Goidenko}}, \bibinfo {author} {\bibfnamefont {I.}~\bibnamefont {Tupitsyn}},
  \ and\ \bibinfo {author} {\bibfnamefont {G.}~\bibnamefont {Plunien}},\
  }\href@noop {} {\bibfield  {journal} {\bibinfo  {journal} {Eur.~Phys.~J.~D}\
  }\textbf {\bibinfo {volume} {45}},\ \bibinfo {pages} {171–177} (\bibinfo
  {year} {2007})}\BibitemShut {NoStop}%
\bibitem [{\citenamefont {Jungmann}(2013{\natexlab{b}})}]{Jungmann:2013}%
  \BibitemOpen
  \bibfield  {author} {\bibinfo {author} {\bibfnamefont {K.}~\bibnamefont
  {Jungmann}},\ }\href@noop {} {\bibfield  {journal} {\bibinfo  {journal}
  {Ann.~Phys.}\ }\textbf {\bibinfo {volume} {525}},\ \bibinfo {pages}
  {550–564} (\bibinfo {year} {2013}{\natexlab{b}})}\BibitemShut {NoStop}%
\bibitem [{\citenamefont {Griffith}\ \emph
  {et~al.}(2009{\natexlab{b}})\citenamefont {Griffith}, \citenamefont
  {Swallows}, \citenamefont {Loftus}, \citenamefont {Romalis}, \citenamefont
  {Heckel},\ and\ \citenamefont {Fortson}}]{edm-199-Hg-Griffith-2009}%
  \BibitemOpen
  \bibfield  {author} {\bibinfo {author} {\bibfnamefont {W.~C.}\ \bibnamefont
  {Griffith}}, \bibinfo {author} {\bibfnamefont {M.~D.}\ \bibnamefont
  {Swallows}}, \bibinfo {author} {\bibfnamefont {T.~H.}\ \bibnamefont
  {Loftus}}, \bibinfo {author} {\bibfnamefont {M.~V.}\ \bibnamefont {Romalis}},
  \bibinfo {author} {\bibfnamefont {B.~R.}\ \bibnamefont {Heckel}}, \ and\
  \bibinfo {author} {\bibfnamefont {E.~N.}\ \bibnamefont {Fortson}},\
  }\href@noop {} {\bibfield  {journal} {\bibinfo  {journal} {Phys.~Rev.~Lett.}\
  }\textbf {\bibinfo {volume} {102}},\ \bibinfo {pages} {101601} (\bibinfo
  {year} {2009}{\natexlab{b}})}\BibitemShut {NoStop}%
\bibitem [{\citenamefont {Amini}\ \emph {et~al.}(2007)\citenamefont {Amini},
  \citenamefont {{Munger,~Jr.}},\ and\ \citenamefont {Gould}}]{Gould:2007}%
  \BibitemOpen
  \bibfield  {author} {\bibinfo {author} {\bibfnamefont {J.~M.}\ \bibnamefont
  {Amini}}, \bibinfo {author} {\bibfnamefont {C.~T.}\ \bibnamefont
  {{Munger,~Jr.}}}, \ and\ \bibinfo {author} {\bibfnamefont {H.}~\bibnamefont
  {Gould}},\ }\href@noop {} {\bibfield  {journal} {\bibinfo  {journal}
  {Phys.~Rev.~A}\ }\textbf {\bibinfo {volume} {75}},\ \bibinfo {pages} {063416}
  (\bibinfo {year} {2007})}\BibitemShut {NoStop}%
\bibitem [{\citenamefont {Romalis}\ \emph {et~al.}(2001)\citenamefont
  {Romalis}, \citenamefont {Griffith}, \citenamefont {Jacobs},\ and\
  \citenamefont {Fortson}}]{edm-199-Hg-Romalis-2001}%
  \BibitemOpen
  \bibfield  {author} {\bibinfo {author} {\bibfnamefont {M.~V.}\ \bibnamefont
  {Romalis}}, \bibinfo {author} {\bibfnamefont {W.~C.}\ \bibnamefont
  {Griffith}}, \bibinfo {author} {\bibfnamefont {J.~P.}\ \bibnamefont
  {Jacobs}}, \ and\ \bibinfo {author} {\bibfnamefont {E.~N.}\ \bibnamefont
  {Fortson}},\ }\href@noop {} {\bibfield  {journal} {\bibinfo  {journal}
  {Phys.~Rev.~Lett.}\ }\textbf {\bibinfo {volume} {86}},\ \bibinfo {pages}
  {2505} (\bibinfo {year} {2001})}\BibitemShut {NoStop}%
\bibitem [{\citenamefont {Swallows}\ \emph
  {et~al.}(2013{\natexlab{b}})\citenamefont {Swallows}, \citenamefont {Loftus},
  \citenamefont {Griffith}, \citenamefont {Heckel},\ and\ \citenamefont
  {Fortson}}]{edm-199-Hg-Swallows:2013}%
  \BibitemOpen
  \bibfield  {author} {\bibinfo {author} {\bibfnamefont {M.~D.}\ \bibnamefont
  {Swallows}}, \bibinfo {author} {\bibfnamefont {T.~H.}\ \bibnamefont
  {Loftus}}, \bibinfo {author} {\bibfnamefont {W.~C.}\ \bibnamefont
  {Griffith}}, \bibinfo {author} {\bibfnamefont {B.~R.}\ \bibnamefont
  {Heckel}}, \ and\ \bibinfo {author} {\bibfnamefont {E.~N.}\ \bibnamefont
  {Fortson}},\ }\href@noop {} {\bibfield  {journal} {\bibinfo  {journal}
  {Phys.~Rev.~A}\ }\textbf {\bibinfo {volume} {87}},\ \bibinfo {pages} {012102}
  (\bibinfo {year} {2013}{\natexlab{b}})}\BibitemShut {NoStop}%
\bibitem [{\citenamefont {Wieman}\ \emph {et~al.}(1999)\citenamefont {Wieman},
  \citenamefont {Pritchard},\ and\ \citenamefont
  {Wineland}}]{Wieman:trapping:1999}%
  \BibitemOpen
  \bibfield  {author} {\bibinfo {author} {\bibfnamefont {C.~E.}\ \bibnamefont
  {Wieman}}, \bibinfo {author} {\bibfnamefont {D.~E.}\ \bibnamefont
  {Pritchard}}, \ and\ \bibinfo {author} {\bibfnamefont {D.~J.}\ \bibnamefont
  {Wineland}},\ }\href@noop {} {\bibfield  {journal} {\bibinfo  {journal}
  {Rev.~Mod.~Phys.}\ }\textbf {\bibinfo {volume} {71}},\ \bibinfo {pages}
  {S253} (\bibinfo {year} {1999})}\BibitemShut {NoStop}%
\bibitem [{\citenamefont {De}\ \emph {et~al.}(2015)\citenamefont {De},
  \citenamefont {Dammalapati},\ and\ \citenamefont
  {Willmann}}]{Willmann:barium:2015}%
  \BibitemOpen
  \bibfield  {author} {\bibinfo {author} {\bibfnamefont {S.}~\bibnamefont
  {De}}, \bibinfo {author} {\bibfnamefont {U.}~\bibnamefont {Dammalapati}}, \
  and\ \bibinfo {author} {\bibfnamefont {L.}~\bibnamefont {Willmann}},\
  }\href@noop {} {\bibfield  {journal} {\bibinfo  {journal} {Phys.~Rev.~A}\
  }\textbf {\bibinfo {volume} {91}},\ \bibinfo {pages} {032517} (\bibinfo
  {year} {2015})}\BibitemShut {NoStop}%
\bibitem [{\citenamefont {Wineland}(2013)}]{Wineland:nobellecture:2013}%
  \BibitemOpen
  \bibfield  {author} {\bibinfo {author} {\bibfnamefont {D.~J.}\ \bibnamefont
  {Wineland}},\ }\href@noop {} {\bibfield  {journal} {\bibinfo  {journal}
  {Rev.~Mod.~Phys.}\ }\textbf {\bibinfo {volume} {85}},\ \bibinfo {pages}
  {1103} (\bibinfo {year} {2013})}\BibitemShut {NoStop}%
\bibitem [{\citenamefont {Ludlow}\ \emph {et~al.}(2015)\citenamefont {Ludlow},
  \citenamefont {Boyd},\ and\ \citenamefont {Ye}}]{Ludlow:clocks:2015}%
  \BibitemOpen
  \bibfield  {author} {\bibinfo {author} {\bibfnamefont {A.~D.}\ \bibnamefont
  {Ludlow}}, \bibinfo {author} {\bibfnamefont {M.~M.}\ \bibnamefont {Boyd}}, \
  and\ \bibinfo {author} {\bibfnamefont {J.}~\bibnamefont {Ye}},\ }\href@noop
  {} {\bibfield  {journal} {\bibinfo  {journal} {Rev.~Mod.~Phys.}\ }\textbf
  {\bibinfo {volume} {87}},\ \bibinfo {pages} {637} (\bibinfo {year}
  {2015})}\BibitemShut {NoStop}%
\bibitem [{\citenamefont {Parker}\ \emph {et~al.}(2015)\citenamefont {Parker},
  \citenamefont {Dietrich}, \citenamefont {Kalita}, \citenamefont {Lemke},
  \citenamefont {Bailey}, \citenamefont {Bishof}, \citenamefont {Greene},
  \citenamefont {Holt}, \citenamefont {Korsch}, \citenamefont {Lu},
  \citenamefont {Mueller}, \citenamefont {O'Connor},\ and\ \citenamefont
  {Singh}}]{Lu:edm-Ra:2015}%
  \BibitemOpen
  \bibfield  {author} {\bibinfo {author} {\bibfnamefont {R.~H.}\ \bibnamefont
  {Parker}}, \bibinfo {author} {\bibfnamefont {M.~R.}\ \bibnamefont
  {Dietrich}}, \bibinfo {author} {\bibfnamefont {M.~R.}\ \bibnamefont
  {Kalita}}, \bibinfo {author} {\bibfnamefont {N.~D.}\ \bibnamefont {Lemke}},
  \bibinfo {author} {\bibfnamefont {K.~G.}\ \bibnamefont {Bailey}}, \bibinfo
  {author} {\bibfnamefont {M.~N.}\ \bibnamefont {Bishof}}, \bibinfo {author}
  {\bibfnamefont {J.~P.}\ \bibnamefont {Greene}}, \bibinfo {author}
  {\bibfnamefont {R.~J.}\ \bibnamefont {Holt}}, \bibinfo {author}
  {\bibfnamefont {W.}~\bibnamefont {Korsch}}, \bibinfo {author} {\bibfnamefont
  {Z.-T.}\ \bibnamefont {Lu}}, \bibinfo {author} {\bibfnamefont
  {P.}~\bibnamefont {Mueller}}, \bibinfo {author} {\bibfnamefont {T.~P.}\
  \bibnamefont {O'Connor}}, \ and\ \bibinfo {author} {\bibfnamefont {J.~T.}\
  \bibnamefont {Singh}},\ }\href@noop {} {\bibfield  {journal} {\bibinfo
  {journal} {Phys.~Rev.~Lett.}\ }\textbf {\bibinfo {volume} {114}},\ \bibinfo
  {pages} {233002} (\bibinfo {year} {2015})}\BibitemShut {NoStop}%
\bibitem [{\citenamefont {Cheal}\ and\ \citenamefont
  {Flanagan}(2010)}]{Cheal:progress:2010}%
  \BibitemOpen
  \bibfield  {author} {\bibinfo {author} {\bibfnamefont {B.}~\bibnamefont
  {Cheal}}\ and\ \bibinfo {author} {\bibfnamefont {K.~T.}\ \bibnamefont
  {Flanagan}},\ }\href@noop {} {\bibfield  {journal} {\bibinfo  {journal}
  {J.~Phys.~G:~Nucl.~Part.~Phys.}\ }\textbf {\bibinfo {volume} {37}},\ \bibinfo
  {pages} {113101} (\bibinfo {year} {2010})}\BibitemShut {NoStop}%
\bibitem [{\citenamefont {Oganessian}(2012)}]{Oganessian:island:2012}%
  \BibitemOpen
  \bibfield  {author} {\bibinfo {author} {\bibfnamefont {Y.}~\bibnamefont
  {Oganessian}},\ }\href@noop {} {\bibfield  {journal} {\bibinfo  {journal}
  {J.~Phys.:~Conf.~Ser.}\ }\textbf {\bibinfo {volume} {337}},\ \bibinfo {pages}
  {012005} (\bibinfo {year} {2012})}\BibitemShut {NoStop}%
\bibitem [{\citenamefont {Sasmal}\ \emph {et~al.}(2016)\citenamefont {Sasmal},
  \citenamefont {Pathak}, \citenamefont {Nayak}, \citenamefont {Vaval},\ and\
  \citenamefont {Pal}}]{Sasmal:2016}%
  \BibitemOpen
  \bibfield  {author} {\bibinfo {author} {\bibfnamefont {S.}~\bibnamefont
  {Sasmal}}, \bibinfo {author} {\bibfnamefont {H.}~\bibnamefont {Pathak}},
  \bibinfo {author} {\bibfnamefont {M.~K.}\ \bibnamefont {Nayak}}, \bibinfo
  {author} {\bibfnamefont {N.}~\bibnamefont {Vaval}}, \ and\ \bibinfo {author}
  {\bibfnamefont {S.}~\bibnamefont {Pal}},\ }\href@noop {} {\bibfield
  {journal} {\bibinfo  {journal} {The Journal of Chemical Physics}\ }\textbf
  {\bibinfo {volume} {144}},\ \bibinfo {pages} {124307} (\bibinfo {year}
  {2016})}\BibitemShut {NoStop}%
\bibitem [{\citenamefont {Norrgard}\ \emph {et~al.}(2016)\citenamefont
  {Norrgard}, \citenamefont {McCarron}, \citenamefont {Steinecker},
  \citenamefont {Tarbutt},\ and\ \citenamefont
  {DeMille}}]{Norrgard-DeMille:2016}%
  \BibitemOpen
  \bibfield  {author} {\bibinfo {author} {\bibfnamefont {E.~B.}\ \bibnamefont
  {Norrgard}}, \bibinfo {author} {\bibfnamefont {D.~J.}\ \bibnamefont
  {McCarron}}, \bibinfo {author} {\bibfnamefont {M.~H.}\ \bibnamefont
  {Steinecker}}, \bibinfo {author} {\bibfnamefont {M.~R.}\ \bibnamefont
  {Tarbutt}}, \ and\ \bibinfo {author} {\bibfnamefont {D.}~\bibnamefont
  {DeMille}},\ }\href@noop {} {\bibfield  {journal} {\bibinfo  {journal}
  {Phys.~Rev.~Lett.}\ }\textbf {\bibinfo {volume} {116}},\ \bibinfo {pages}
  {063004} (\bibinfo {year} {2016)})}\BibitemShut {NoStop}%
\bibitem [{\citenamefont {Prehn}\ \emph {et~al.}(2016)\citenamefont {Prehn},
  \citenamefont {Ibr{\"u}gger}, \citenamefont {Gl{\"o}ckner}, \citenamefont
  {Rempe},\ and\ \citenamefont {Zeppenfeld}}]{Prehn:2016}%
  \BibitemOpen
  \bibfield  {author} {\bibinfo {author} {\bibfnamefont {A.}~\bibnamefont
  {Prehn}}, \bibinfo {author} {\bibfnamefont {M.}~\bibnamefont {Ibr{\"u}gger}},
  \bibinfo {author} {\bibfnamefont {R.}~\bibnamefont {Gl{\"o}ckner}}, \bibinfo
  {author} {\bibfnamefont {G.}~\bibnamefont {Rempe}}, \ and\ \bibinfo {author}
  {\bibfnamefont {M.}~\bibnamefont {Zeppenfeld}},\ }\href@noop {} {\bibfield
  {journal} {\bibinfo  {journal} {Phys.~Rev.~Lett.}\ }\textbf {\bibinfo
  {volume} {116}},\ \bibinfo {pages} {063005} (\bibinfo {year}
  {2016})}\BibitemShut {NoStop}%
\bibitem [{\citenamefont {Isaev}\ and\ \citenamefont
  {Berger}(2016)}]{Isaev-Berger:2016}%
  \BibitemOpen
  \bibfield  {author} {\bibinfo {author} {\bibfnamefont {T.~A.}\ \bibnamefont
  {Isaev}}\ and\ \bibinfo {author} {\bibfnamefont {R.}~\bibnamefont {Berger}},\
  }\href@noop {} {\bibfield  {journal} {\bibinfo  {journal} {Phys.~Rev.~Lett.}\
  }\textbf {\bibinfo {volume} {116}},\ \bibinfo {pages} {063006} (\bibinfo
  {year} {2016})}\BibitemShut {NoStop}%
\bibitem [{\citenamefont {Panda}\ \emph {et~al.}(2016)\citenamefont {Panda},
  \citenamefont {O’Leary}, \citenamefont {West}, \citenamefont {Baron},
  \citenamefont {Hess}, \citenamefont {Hoffman}, \citenamefont {Kirilov},
  \citenamefont {Overstreet}, \citenamefont {West}, \citenamefont {DeMille},
  \citenamefont {Doyle},\ and\ \citenamefont
  {Gabrielse}}]{Panda-Gabrielse:2016}%
  \BibitemOpen
  \bibfield  {author} {\bibinfo {author} {\bibfnamefont {C.~D.}\ \bibnamefont
  {Panda}}, \bibinfo {author} {\bibfnamefont {B.~R.}\ \bibnamefont
  {O’Leary}}, \bibinfo {author} {\bibfnamefont {A.~D.}\ \bibnamefont {West}},
  \bibinfo {author} {\bibfnamefont {J.}~\bibnamefont {Baron}}, \bibinfo
  {author} {\bibfnamefont {P.~W.}\ \bibnamefont {Hess}}, \bibinfo {author}
  {\bibfnamefont {C.}~\bibnamefont {Hoffman}}, \bibinfo {author} {\bibfnamefont
  {E.}~\bibnamefont {Kirilov}}, \bibinfo {author} {\bibfnamefont {C.~B.}\
  \bibnamefont {Overstreet}}, \bibinfo {author} {\bibfnamefont {E.~P.}\
  \bibnamefont {West}}, \bibinfo {author} {\bibfnamefont {D.}~\bibnamefont
  {DeMille}}, \bibinfo {author} {\bibfnamefont {J.~M.}\ \bibnamefont {Doyle}},
  \ and\ \bibinfo {author} {\bibfnamefont {G.}~\bibnamefont {Gabrielse}},\
  }\href@noop {} {\  (\bibinfo {year} {2016})},\ \Eprint
  {http://arxiv.org/abs/1603.07707v1} {arXiv:1603.07707v1 [physics.atom-ph]}
  \BibitemShut {NoStop}%
\bibitem [{\citenamefont {Fleig}\ \emph {et~al.}(2016)\citenamefont {Fleig},
  \citenamefont {Nayak},\ and\ \citenamefont {Kozlov}}]{Fleig:2016}%
  \BibitemOpen
  \bibfield  {author} {\bibinfo {author} {\bibfnamefont {T.}~\bibnamefont
  {Fleig}}, \bibinfo {author} {\bibfnamefont {M.~K.}\ \bibnamefont {Nayak}}, \
  and\ \bibinfo {author} {\bibfnamefont {M.~G.}\ \bibnamefont {Kozlov}},\
  }\href@noop {} {\bibfield  {journal} {\bibinfo  {journal} {Phys.~Rev.~A}\
  }\textbf {\bibinfo {volume} {93}},\ \bibinfo {pages} {012505} (\bibinfo
  {year} {2016})}\BibitemShut {NoStop}%
\bibitem [{\citenamefont {Leanhardt}\ \emph {et~al.}(2011)\citenamefont
  {Leanhardt}, \citenamefont {Bohn}, \citenamefont {Loh}, \citenamefont
  {Maletinsky}, \citenamefont {Meyer}, \citenamefont {Sinclair}, \citenamefont
  {Stutz},\ and\ \citenamefont {Cornell}}]{Leanhardt-Cornell:2011}%
  \BibitemOpen
  \bibfield  {author} {\bibinfo {author} {\bibfnamefont {A.~E.}\ \bibnamefont
  {Leanhardt}}, \bibinfo {author} {\bibfnamefont {J.~L.}\ \bibnamefont {Bohn}},
  \bibinfo {author} {\bibfnamefont {H.}~\bibnamefont {Loh}}, \bibinfo {author}
  {\bibfnamefont {P.}~\bibnamefont {Maletinsky}}, \bibinfo {author}
  {\bibfnamefont {E.~R.}\ \bibnamefont {Meyer}}, \bibinfo {author}
  {\bibfnamefont {L.~C.}\ \bibnamefont {Sinclair}}, \bibinfo {author}
  {\bibfnamefont {R.~P.}\ \bibnamefont {Stutz}}, \ and\ \bibinfo {author}
  {\bibfnamefont {E.~A.}\ \bibnamefont {Cornell}},\ }\href@noop {} {\bibfield
  {journal} {\bibinfo  {journal} {J.~Mol.~Spectrosc.}\ }\textbf {\bibinfo
  {volume} {270}},\ \bibinfo {pages} {1} (\bibinfo {year} {2011})}\BibitemShut
  {NoStop}%
\bibitem [{\citenamefont {Gresh}\ \emph {et~al.}(2016)\citenamefont {Gresh},
  \citenamefont {Cossel}, \citenamefont {Zhou}, \citenamefont {Ye},\ and\
  \citenamefont {Cornell}}]{Gresh-Cornell:2016}%
  \BibitemOpen
  \bibfield  {author} {\bibinfo {author} {\bibfnamefont {D.~N.}\ \bibnamefont
  {Gresh}}, \bibinfo {author} {\bibfnamefont {K.~C.}\ \bibnamefont {Cossel}},
  \bibinfo {author} {\bibfnamefont {Y.}~\bibnamefont {Zhou}}, \bibinfo {author}
  {\bibfnamefont {J.}~\bibnamefont {Ye}}, \ and\ \bibinfo {author}
  {\bibfnamefont {E.~A.}\ \bibnamefont {Cornell}},\ }\href@noop {} {\bibfield
  {journal} {\bibinfo  {journal} {J.~Mol.~Spectrosc.}\ }\textbf {\bibinfo
  {volume} {319}},\ \bibinfo {pages} {1} (\bibinfo {year} {2016})}\BibitemShut
  {NoStop}%
\bibitem [{God()}]{Goddard}%
  \BibitemOpen
  \href@noop {} {}\bibinfo {note} {Attributed to Robert H.~Goddard}\BibitemShut
  {NoStop}%
\end{thebibliography}%

\end{document}